{}
{}

\documentclass[12pt,a4paper]{article}

\newif\ifpublic\publictrue

\usepackage{amsmath,amssymb,mathtools}
\ifpublic\else\usepackage{showkeys}\fi
\usepackage[bookmarks=true,hyperfigures=true]{hyperref}
\usepackage[nosort]{cite}
\usepackage[bulletsep]{collref}

\def\showkeysrefformat#1{{\normalfont\tiny\ttfamily#1}}
\makeatletter
\def\SK@@ref#1>#2\SK@{%
 {\@inlabelfalse\leavevmode\vbox to\z@{%
 \vss\SK@refcolor\rlap{\vrule\raise .75em%
  \hbox{\showkeysrefformat{#2}}}}}}
\makeatother

\usepackage[a4paper,text={160mm,247mm},centering]{geometry}

\allowdisplaybreaks[3]

\numberwithin{equation}{section}

\expandafter\def\expandafter\bfseries\expandafter{\bfseries\ifmmode\else\boldmath\fi}
\expandafter\def\expandafter\mdseries\expandafter{\mdseries\ifmmode\else\unboldmath\fi}
\expandafter\def\expandafter\normalfont\expandafter{\normalfont\ifmmode\else\unboldmath\fi}

\RequirePackage{verbatim}

\makeatletter
\newwrite\bibinl@out
\newenvironment{bibtex}[1][\jobname]{%
  \immediate\openout\bibinl@out #1.bib
  \immediate\write\bibinl@out{\@percentchar generated from `\jobname' starting line \the\inputlineno^^J}%
  \def\verbatim@processline{\immediate\write\bibinl@out{\the\verbatim@line}}%
  \@bsphack\let\do\@makeother\dospecials\catcode`\^^M\active\verbatim@start
}%
{\immediate\closeout\bibinl@out\@esphack}
\makeatother

\usepackage{fixmath}

\let\barefrac=\frac
\renewcommand{\frac}[2]{\mathinner{\barefrac{#1}{#2}}}

\let\baresqrt=\sqrt
\makeatletter
\renewcommand{\sqrt}{\@ifnextchar[\@sqrt@space@a\@sqrt@space@b}
\def\@sqrt@space@a[#1]#2{\mathinner{\mathchoice{\mkern-3mu}{\mkern-3mu}{}{}\baresqrt[#1]{#2}}}
\def\@sqrt@space@b#1{\mathinner{\mathchoice{\mkern-3mu}{\mkern-3mu}{}{}\baresqrt{#1}}}
\makeatother

\makeatletter
\let\per@dot@old=\.
\def\.{\ifmmode\def\per@dot@sel{\mkern3mu}\else\def\per@dot@sel{\per@dot@old}\fi\per@dot@sel}
\makeatother

\newcommand\eqq{\vcentcolon=}

\newcommand{\sfrac}[2]{{\textstyle\frac{#1}{#2}}}
\newcommand{\half}{\sfrac{1}{2}}

\newcommand{\vfrac}[2]{\ifmmode\mathinner{\textstyle^{#1}\!/\!_{#2}}\else$^{#1}\!/\!_{#2}$\fi}

\newcommand{\Real}{\mathbb{R}}
\newcommand{\Complex}{\mathbb{C}}

\newcommand{\Natural}{\mathbb{N}}

\newcommand{\hopf}[1]{\mathrm{#1}}
\newcommand{\env}{\hopf{U}}
\newcommand{\kap}{\hopf{K}}

\newcommand{\alg}[1]{\mathfrak{#1}}

\newcommand{\copro}{\mathrm{\Delta}}
\newcommand{\cobrack}{\delta}
\newcommand{\cop}{\text{cop}}

\newcommand{\directsum}{\times}

\newcommand{\aut}{{\scriptscriptstyle\text{A}}}
\newcommand{\ctr}{{\scriptscriptstyle\text{C}}}
\newcommand{\comp}{{\scriptscriptstyle\text{B}}}
\newcommand{\rot}{T}

\newcommand{\antipode}{\mathrm{S}}

\newcommand{\rmat}{\mathcal{R}}
\newcommand{\Casimir}{x}

\usepackage[extdef]{delimset}

\DeclareMathOperator{\diag}{diag}

\DeclareMathOperator{\Li}{Li}
\DeclareMathOperator{\Ad}{Ad}
\DeclareMathOperator{\ad}{ad}
\newcommand{\Order}{\mathcal{O}}

\def\[{\begin{equation}}
\def\]{\end{equation}}
\newcommand{\nln}{\nonumber\\}

\providecommand{\href}[2]{#2}
\newcommand{\arxivlink}[1]{\href{http://arxiv.org/abs/#1}{arxiv:#1}}

\makeatletter
\def\mr@ignsp#1 {\ifx\:#1\@empty\else #1\expandafter\mr@ignsp\fi}%
\newcommand{\multiref}[1]{\begingroup
  \xdef\mr@no@sparg{\expandafter\mr@ignsp#1 \: }%
  \def\mr@comma{}\def\mr@dash{-}%
  \@for\mr@refs:=\mr@no@sparg\do{%
    \ifx\mr@refs\mr@dash\def\mr@comma{}--\else%
    \mr@comma\def\mr@comma{,}\ref{\mr@refs}\fi}%
\endgroup}
\renewcommand{\eqref}[1]{(\multiref{#1})}
\makeatother

\makeatletter
\newcommand{\namedref}[2]{\hyperref[#2]{#1~\ref*{#2}}}
\newcommand{\secref}{\@ifstar{\namedref{Section}}{\namedref{sec.}}}
\newcommand{\appref}{\@ifstar{\namedref{Appendix}}{\namedref{app.}}}
\newcommand{\tabref}{\@ifstar{\namedref{Table}}{\namedref{tab.}}}
\newcommand{\figref}{\@ifstar{\namedref{Figure}}{\namedref{fig.}}}
\makeatother

\let\oldbib=\thebibliography
\def\thebibliography{\phantomsection\addcontentsline{toc}{section}{\refname}\oldbib}

\let\oldtoc=\tableofcontents
\def\tableofcontents{\phantomsection\addcontentsline{toc}{section}{\contentsname}\oldtoc}

\providecommand{\hypersetup}[1]{}
\providecommand{\texorpdfstring}[2]{#1}

\hypersetup{plainpages=false}
\hypersetup{pdfpagemode=UseNone}
\hypersetup{bookmarksnumbered=true}
\hypersetup{pdfstartview=FitH}
\hypersetup{colorlinks=false}
\hypersetup{citebordercolor={.5 1 .5}}
\hypersetup{urlbordercolor={.5 1 1}}
\hypersetup{linkbordercolor={1 .7 .7}}

\makeatletter
\let\@keywords\@empty
\let\@subject\@empty
\providecommand{\keywords}[1]{\gdef\@keywords{#1}}
\providecommand{\subject}[1]{\gdef\@subject{#1}}
\def\thetitle{\@title}
\def\theauthor{\@author}
\def\thesubject{\@subject}
\def\thedate{\@date}
\def\thekeywords{\@keywords}
\makeatother
\AtBeginDocument{
\hypersetup{pdftitle={\thetitle}}%
\hypersetup{pdfauthor={\theauthor}}%
\hypersetup{pdfsubject={\thesubject}}%
\hypersetup{pdfkeywords={\thekeywords}}%
}

\newif\ifshownote
\shownotetrue

\ifpublic\shownotefalse\fi

\ifshownote

\ifpdf\else\RequirePackage[active]{srcltx}\fi
\RequirePackage{color}
\RequirePackage{ulem}

\newcommand{\remark}[2][]{{\normalfont\sffamily\hspace{1ex}%
  \def\emph{\textsl}\def\textbullet{$\bullet$}
  \def\tmparga{#1}%
  \def\tmpargb{RH}\ifx\tmparga\tmpargb\color[rgb]{0,0,0.8}\fi%
  \def\tmpargb{NB}\ifx\tmparga\tmpargb\color[rgb]{0,0.6,0}\fi%
  \def\tmpargb{BH}\ifx\tmparga\tmpargb\color[rgb]{0.7,0,0}\fi%
  \def\tmpargb{}\ifx\tmparga\tmpargb\color{red}\fi%
  \def\tmpargb{}\ifx\tmparga\tmpargb\else \textbf{#1:} \fi%
  #2\hspace{1ex}}}

\else
\newcommand{\remark}[2][]{\ignorespaces}

\fi

\title{Maximally extended \texorpdfstring{$\alg{sl}(2|2)$}{sl(2|2)},
\texorpdfstring{\unskip\\}{}%
q-deformed \texorpdfstring{$\alg d(2,1;\epsilon)$}{d(2,1;eps)}
\texorpdfstring{\unskip\\}{}%
and 3D kappa-Poincar\'e}

\begin{document}

\pdfbookmark[1]{Title Page}{title}
\thispagestyle{empty}

\begingroup\raggedleft\footnotesize\ttfamily
\arxivlink{1704.05093}
\par\endgroup

\vspace*{2cm}
\begin{center}%
\begingroup\Large\bfseries\thetitle\par\endgroup
\vspace{1cm}


\begingroup\scshape
Niklas Beisert,
Reimar Hecht and
Ben Hoare
\endgroup
\vspace{5mm}

\textit{Institut f\"ur Theoretische Physik,\\
Eidgen\"ossische Technische Hochschule Z\"urich,\\
Wolfgang-Pauli-Strasse 27, 8093 Z\"urich, Switzerland}
\vspace{0.1cm}

\begingroup\ttfamily\small
\verb+{+nbeisert,hechtr,bhoare\verb+}+@itp.phys.ethz.ch\par
\endgroup
\vspace{5mm}

\vfill

\textbf{Abstract}\vspace{5mm}

\begin{minipage}{12.7cm}
We show that the maximal extension 
$\alg{sl}(2)\ltimes\alg{psl}(2|2)\ltimes \Complex^3$
of the $\alg{sl}(2|2)$ superalgebra 
can be obtained as a contraction limit of the semi-simple superalgebra
$\alg{d}(2,1;\epsilon) \directsum \alg{sl}(2)$.
We reproduce earlier results on the corresponding q-deformed Hopf algebra 
and its universal R-matrix by means of contraction.
We make the curious observation that 
the above algebra is related to kappa-Poincar\'e symmetry.
When dropping the graded part $\alg{psl}(2|2)$
we find a novel one-parameter deformation of the 3D kappa-Poincar\'e algebra.
Our construction also provides a concise exact expression for its universal R-matrix.
\end{minipage}

\vspace*{4cm}

In Honour of Petr P.\ Kulish and Ludvig D.\ Faddeev

\end{center}

\newpage

\section{Introduction}

During the 1970's and 80's the Leningrad school led by Ludvig Faddeev
developed the quantum inverse scattering method
and the algebraic Bethe ansatz
to solve large classes of quantum integrable systems
\cite{Faddeev:1959yc,Faddeev:1979gh}, see also \cite{Faddeev:1996iy}.
These developments laid the foundations for the mathematical formulation
of quantum groups and algebras
\cite{Drinfeld:1985rx,Drinfeld:1986in,Jimbo:1985zk,Jimbo:1985vd}. 
The theory of quantum algebras based on simple Lie algebras 
has since been worked out quite exhaustively  
with numerous results, methods and applications,
see e.g.\ \cite{Faddeev:1987ih,Chari:1994pz}.
These quantum algebras display very regular structures 
which are closely related to their root systems.

Petr Kulish was among the first to formulate 
quantum algebras based on Lie superalgebras
\cite{Kulish:1988gr,Kulish:1989sv}.
The generalisation to simple superalgebras
largely uses the same regular structures associated to their root systems,
and hence many such algebras can be constructed along similar lines.
However, there are also a few special cases with peculiar features.
For example, there are simple Lie superalgebras such as $\alg{psl}(N|N)$ 
with vanishing dual Coxeter number.
Furthermore, there are the exceptional superalgebras $\alg{d}(2,1;\alpha)$,
which form a family depending on the continuous parameter $\alpha$.
While similar features do not arise for ordinary simple Lie algebras,
they may do for non-simple Lie algebras, with the corresponding quantum algebras 
sometimes displaying novel structures and applications.
Consequently, quantum algebras based on non-simple Lie algebras 
or Lie superalgebras are explored to a lesser extent, 
and may still harbour some pleasant surprises.

Examples of such unconventional quantum algebra structures 
have been found in the one-dimensional Hubbard model 
and in $\mathcal{N}=4$ supersymmetric gauge theory in the planar limit
(as well as related models in the context of the AdS/CFT correspondence),
see \cite{HubbBook} and \cite{Beisert:2010jr} for reviews. 
In particular, they both possess a peculiar R-matrix that is not of difference form
\cite{Shastry:1986zz,Beisert:2005tm,Beisert:2006qh}.
Due to its uncommon structure, this R-matrix escapes 
the established classification in terms of Yangian and quantum affine algebras
based on semi-simple Lie (super)algebras.

A long-standing goal in this regard is to construct the 
underlying quantum algebra and its universal R-matrix.
Several pieces of this puzzle are known. 
It is clear that the Lie superalgebra $\alg{sl}(2|2)$ 
and its exceptional central extensions plays a role
\cite{Beisert:2005tm,Gomez:2006va,Plefka:2006ze}.
Studies of the classical limit have demonstrated that 
the complete algebra also needs to be extended by derivations
\cite{Torrielli:2007mc,Matsumoto:2007rh,Beisert:2007ty}.
The maximal extension of $\alg{sl}(2|2)$ 
is a non-simple superalgebra
\unskip\footnote{Throughout the paper we will 
assume algebras to be over the complex numbers.
The choice of signature is relevant only to real forms and 
consequently it will not be of concern to us.}
\[
\alg{sl}(2)\ltimes \alg{psl}(2|2) \ltimes \Complex^3,
\]
which incidentally also serves as a non-standard extended super-Poincar\'e symmetry
in three spacetime dimensions \cite{Nahm:1977tg}.
The classical analysis \cite{Beisert:2007ty} suggests that 
the relevant quantum algebra is a peculiar subalgebra,
yet to be identified,
of the Yangian of the maximal extension of $\alg{sl}(2|2)$.

A complication related to the latter approach is that 
the underlying Lie algebra is non-simple and the precise form 
of its algebra relations does not necessarily follow 
the patterns known from simple Lie (super)algebras.
Moreover, the Yangian is a contraction limit 
of the quantum affine algebra
which obscures some of its uniform structure.
The q-deformation for the maximally extended $\alg{sl}(2|2)$
was explored in \cite{Beisert:2016qei} and revealed
some unconventional terms in the algebra and in the R-matrix.
Unfortunately, the form of the result does not make 
evident how to construct the exact form 
of q-deformed non-simple Lie (super)algebras, 
except by applying some amount of trial and error and brute force.

\bigskip

In this paper we revisit the q-deformation 
of maximally extended $\alg{sl}(2|2)$.
We will use a different method to construct 
the algebra and its universal R-matrix. 
Our idea is based on the connection 
between extensions of $\alg{psl}(2|2)$ 
and the exceptional superalgebra $\alg{d}(2,1;\epsilon)$ for $\epsilon=0$.
In fact, there are two ways the limit $\epsilon\to 0$ can be approached,
and they yield the superalgebra $\alg{psl}(2|2)$,
either with three central extensions $\Complex^3$
or with an $\alg{sl}(2)$ algebra of derivations.
\unskip\footnote{There may be further ways of taking the limit, 
and attempts have been made to obtain the relevant 
algebra for the above R-matrix along these lines \cite{Matsumoto:2008ww}.}
Curiously, both of these extensions can coexist in a consistent 
Lie superalgebra, which is the maximal extension of $\alg{sl}(2|2)$.
However, they cannot both be obtained at the same time 
from $\alg{d}(2,1;\epsilon)$ alone
as the latter lacks three generators.
To overcome this shortcoming, we can supply three more generators
forming an $\alg{sl}(2)$ algebra, 
and indeed there is a contraction limit 
that yields the maximally extended $\alg{sl}(2|2)$ 
\[
\alg{d}(2,1;\epsilon) \directsum \alg{sl}(2)
\quad\stackrel{\epsilon\to0}{\longrightarrow}\quad 
\alg{sl}(2)\ltimes \alg{psl}(2|2) \ltimes \Complex^3.
\]

We will use this contraction limit to construct 
q-deformed maximally extended $\alg{sl}(2|2)$
based on the (standard) q-deformations of $\alg{d}(2,1;\epsilon)$
and $\alg{sl}(2)$.
We will show that this construction yields precisely the algebra relations
and the R-matrix obtained in \cite{Beisert:2016qei}.
The $\alg{d}(2,1;\epsilon)$ origin of maximally extended $\alg{sl}(2|2)$
also explains some of the observed peculiarities.
For instance, the q-deformed $\alg{d}(2,1;\epsilon)$ algebra
has three q-deformed $\alg{sl}(2)$ subalgebras. 
Importantly, these have deformation parameters 
$q$, $q^\epsilon$ and $q^{-1-\epsilon}$, respectively.
This implies that in the limit $\epsilon\to 0$, 
some part of the algebra will be (more or less) undeformed ($q\approx 1$)
while some other parts remain fully deformed ($q\not\approx 1$). 

\medskip

The fully deformed part of the algebra is the superalgebra $\alg{psl}(2|2)$ 
while the $\alg{sl}(2)$ derivations and charges $\Complex^3$ 
are only weakly deformed.
In fact, one can remove the superalgebra $\alg{psl}(2|2)$ 
from the bigger algebra and what remains is a deformation 
of the 3D Poincar\'e algebra
\[\alg{sl}(2)\ltimes \Complex^3=\alg{iso}(3).\]
Deformations of Poincar\'e symmetry and associated physical theories
have been investigated in their own right (see e.g.\ the review articles
\cite{KowalskiGlikman:2004qa,SigmaNonComm,SigmaSpaceTime,Pachol:2011tp})
and this one corresponds to the so-called kappa-Poincar\'e symmetry.
Our algebra turns out to be a novel one-parameter family of
deformations of the kappa-Poincar\'e algebra that is particular to 3D. 
In this sense, q-deformed maximally extended $\alg{sl}(2|2)$
is a supersymmetric extension of 3D kappa-Poincar\'e
(with two deformation parameters).

\bigskip

The present paper is organised as follows:
We start in \secref{sec:sl2+sl2} by performing the contraction 
of the q-deformation of $\alg{so}(4)$ 
to obtain a deformation of the 3D Poincar\'e algebra $\alg{iso}(3)$.
This investigation highlights relevant features in a simplified context, 
which are then used in the generalisation to superalgebras
performed in \secref{sec:sl2+d}.
In \secref{sec:kappoincare} we 
compare our deformation of $\alg{iso}(3)$ 
to the 3D kappa-Poincar\'e algebra
and show that it is a one-parameter deformation of the latter.
We conclude in \secref{sec:concl}
and give an outlook.

\section{Deformation of 3D Poincar\'e as a contraction}
\label{sec:sl2+sl2}

The maximal extension $\alg{sl}(2) \ltimes \alg{psl}(2|2) \ltimes \Complex^3$ 
of the $\alg{sl}(2|2)$ superalgebra 
is a non-standard supersymmetric extension \cite{Nahm:1977tg} 
of the ordinary 3D Poincar\'e algebra 
\[
\alg{iso}(3)=\alg{sl}(2)\ltimes\Complex^3.
\]
The q-deformation of this superalgebra along with its universal R-matrix 
was constructed in \cite{Beisert:2016qei},
and it was seen to possess a number of unusual features. 
Most of these unusual features relate to its 3D Poincar\'e subalgebra,
and in fact they remain present in the restriction to it.
Therefore it makes sense to study the deformed 3D Poincar\'e algebra 
in detail towards understanding the peculiarities in a simpler context.

The 3D Poincar\'e algebra is well-known to be a contraction of 
\[
\alg{so}(4)=\alg{sl}(2)\directsum \alg{sl}(2).
\]
Our goal in this section is to lift this contraction to the q-deformed
algebras.  As we will see, the unusual features alluded to above can all be
understood as originating in this limiting procedure.  It also transpires, as
we will discuss later in \secref{sec:kappoincare}, that the q-deformed algebra
is closely related to kappa-Poincar\'e algebras
\cite{Lukierski:1991pn,Giller:1992xg,Lukierski:1993wxa,Maslanka}.
The latter are understood to be constructible as contractions of
q-deformed orthogonal algebras
\cite{Celeghini:1990bf,Celeghini:1990xx,Lukierski:1991pn,Lukierski:1992dt},
however, the 3D case turns out to be particularly tractable and gives rise to
some special features.  Afterwards, we will return to q-deformed maximally
extended $\alg{sl}(2|2)$ in \secref{sec:sl2+d} and show that it originates
from a contraction involving the exceptional superalgebra
$\alg{d}(2,1;\alpha)$.

\bigskip

In this paper we will be dealing with several q-deformed (sub)algebras 
whose (effective) deformation parameter will take different values $q_i$ 
for some of which we will also take the limit $q_i\to 1$.
In order to parameterise the deformations, 
we will find it convenient to introduce 
a fixed reference deformation parameter
which we denote by $q$ or equivalently by $\hbar$
\[q \equiv e^\hbar.\]
The q-deformations of individual (sub)algebras will be 
specified relative to the reference parameters as
$q^{\alpha}=e^{\alpha\hbar}$.
This will allow us to tune the relative
deformation strength of subalgebras with concrete values, while keeping
$q = e^\hbar$ fixed to specify the overall strength of the deformation.
Furthermore, we shall label q-deformed objects 
such as q-deformed universal enveloping algebras,
q-numbers and q-exponents,
by an index `$\hbar$'
\begin{align}
\env_\hbar(\alg{g})
,\qquad
[n]_\hbar
,\qquad
\exp_\hbar(x)
,\qquad
\ldots
\end{align}
rather than the somewhat more common notation
$\env_q(\alg{g})$, $[n]_q$, $\exp_q(x)$, \ldots.
This helps us specify the relative 
deformation strength as in $\env_{\alpha\hbar}(\alg{g})$
in a slightly more legible fashion.

\subsection{Hopf algebra}

We start by considering two mutually commuting copies of the q-deformed 
$\alg{sl}(2)$ algebra with different deformation parameters
\[
\env_{\epsilon\hbar}(\alg{sl}(2))\otimes \env_{\tilde{\epsilon}\hbar}(\alg{sl}(2)).
\]
For the purpose of taking a contraction limit we introduce 
the two relative deformation parameters $\epsilon$ and $\tilde \epsilon$
which will later be taken to zero.
The first copy $\env_{\epsilon\hbar}(\alg{sl}(2))$ of the algebra
has the following set of defining relations
\begin{align}
\label{eq:sl2q-relation-e}
[H,E] & =2E
,&
\copro E & =E\otimes 1+q^{-\epsilon H}\otimes E
,\\
\label{eq:sl2q-relation-f}
[H,F] & =-2F 
,& \copro F & =F\otimes q^{\epsilon H}+1\otimes F
,\\
\label{eq:sl2q-relation-h}
 [E,F] & =\frac{q^{\epsilon H}-q^{-\epsilon H}}{q^{\epsilon}-q^{-\epsilon}}
,&\copro H & =H\otimes1+1\otimes H.
\end{align}
The second copy $\env_{\tilde \epsilon\hbar}(\alg{sl}(2))$ of the algebra
obeys the same set of defining relations \eqref{eq:sl2q-relation-e,-,eq:sl2q-relation-h}
with the generators $E,F,H$ and parameter $\epsilon$ 
replaced by $\tilde E,\tilde F,\tilde H$ and $\tilde \epsilon$, respectively.

\paragraph{Contraction.}

We now want to perform the contraction limit 
$\alg{sl}(2)\directsum\alg{sl}(2)\to \alg{sl}(2)\ltimes \Complex^3$.
At the same time we also take the limit $\epsilon,\tilde\epsilon\to 0$
which ordinarily removes the q-deformation. 
As usual, the overall limit depends crucially on how the 
various limits are taken relative to each other,
and only for an appropriate fine-tuning of limits
we will find the desired algebra which 
carries some remnants of the q-deformation.

For the contraction limit, it makes sense to introduce 
the following combinations of generators
that we assume to be finite in the limit $\epsilon,\tilde\epsilon\to 0$ 
\footnote{Note that we use the parameter $\epsilon$
not only for performing the contraction 
but also to specify the relative strength of the q-deformation.
This imposes no restriction because there is still the 
overall deformation parameter $\hbar$ which can be adjusted independently.}
\begin{align}
\label{eq:sl(2)+sl(2)DefLP}
E_\aut & \eqq E+\tilde{E}, 
& E_\ctr & \eqq \epsilon E,
\\
F_\aut & \eqq F+\tilde{F}, 
& F_\ctr & \eqq \epsilon F,
\\
H_\aut & \eqq H+\tilde{H}, 
&  H_\ctr & \eqq \epsilon H.
\label{eq:sl(2)+sl(2)DefHC}
\end{align}

Furthermore, it is crucial to take the limit $\epsilon,\tilde\epsilon\to0$ 
in both algebras simultaneously in a coordinated fashion. 
To understand the requirements, let us inspect some algebra relations 
in the new basis \eqref{eq:sl(2)+sl(2)DefLP,-,eq:sl(2)+sl(2)DefHC}.
The commutation relation $[E_\aut,F_\aut]$ takes the form
\begin{align}
[E_\aut,F_\aut]
&=
\frac{q^{ H_\ctr}-q^{- H_\ctr}}{q^{\epsilon}-q^{-\epsilon}}
+\frac{q^{\tilde{\epsilon}H_\aut-(\tilde{\epsilon}/\epsilon)  H_\ctr}
  -q^{-\tilde{\epsilon}H_\aut+(\tilde{\epsilon}/\epsilon)  H_\ctr}}
  {q^{\tilde{\epsilon}}-q^{-\tilde{\epsilon}}}
\nln &=
\frac{1}{\epsilon}
\brk[s]*{\frac{q^{ H_\ctr}-q^{- H_\ctr}}{2\hbar}
+\frac{q^{-\beta  H_\ctr}-q^{\beta  H_\ctr}}{2\beta\hbar}}
+\Order(\epsilon^0),
\end{align}
where the expansion in the second line is based on the assumption that 
$\tilde \epsilon\simeq \beta\epsilon$ in the limit $\epsilon\to 0$.
We see that the divergent term vanishes for the choices $\beta=\pm 1$.
This choice also leads to a well-defined contraction limit 
for all the other algebra relations.
Next we inspect the coalgebra.
The coproduct of $E_\aut$ reads
\begin{align}
\copro E_\aut&=E_\aut\otimes1+q^{-\tilde{\epsilon}H_\aut+
(\tilde{\epsilon}/\epsilon) H_\ctr}\otimes E_\aut
-\frac{1}{\epsilon}
\brk!{q^{-\tilde{\epsilon}H_\aut+(\tilde{\epsilon}/\epsilon) H_\ctr}-q^{- H_\ctr}}\otimes E_\ctr
\nln &=
-\frac{2\hbar}{\epsilon}\brk!{q^{\beta  H_\ctr}-q^{- H_\ctr}}\otimes E_\ctr
+\Order(\epsilon^0),
\end{align}
Evidently, we need to set $\beta=-1$ 
to eliminate the divergent term in this relation.
The same choice will eliminate a similar divergence in $\copro F_\aut$.
Altogether we find that the Hopf algebra has a well-defined 
contraction limit if $\tilde\epsilon/\epsilon\to -1$.
\unskip\footnote{Due to the Hopf algebra isomorphism between 
$\env_{\pm\hbar}(\alg{sl}(2))$
we could alternatively assume $\tilde\epsilon/\epsilon\to +1$
and replace the generators $(\tilde E,\tilde F,\tilde H)$ 
in the basis \eqref{eq:sl(2)+sl(2)DefLP,-,eq:sl(2)+sl(2)DefHC}
by $(q^{-\tilde\epsilon H}\tilde F,\tilde Eq^{\tilde\epsilon H},-\tilde H)$.}

\paragraph{Limit.}

We now write $\tilde{\epsilon}$ as a general expansion in terms of $\epsilon$
subject to the constraint derived above
\[\label{eq:asymlimit}
\tilde{\epsilon}(\epsilon)=-\epsilon+\xi\epsilon^{2}+\Order(\epsilon^{3}).
\]
The parameter $\xi$ can be adjusted freely, and it turns out to survive 
in the limit. 
Higher-order terms in the relationship between $\tilde\epsilon$ and $\epsilon$ 
do not contribute in the limit. 

The limit of the commutation relations reads
\[
\label{eq:bosalg0}
 \comm{E_\aut}{F_\aut} 
= \frac{1}{2}\brk!{q^{  H_\ctr}+q^{ - H_\ctr}}(H_\aut+ \xi H_\ctr)
- \frac{\xi}{2\hbar}\brk!{q^{  H_\ctr}-q^{ - H_\ctr}},
\]
as well as
\begin{align}
\label{eq:bosalg1}
\comm{H_\aut}{E_\aut} & =2E_\aut, 
& \comm{H_\aut}{F_\aut} & =-2F_\aut, 
\\{} 
\label{eq:bosalg2}
[E_\aut,E_\ctr] & =0, 
& [E_\aut,F_\ctr] & =\frac{q^{ H_\ctr}-q^{- H_\ctr}}{2\hbar},
& [E_\aut,H_\ctr] & =-2E_\ctr, 
\\{} 
\label{eq:bosalg3}
[F_\aut,E_\ctr] & =-\frac{q^{ H_\ctr}-q^{- H_\ctr}}{2\hbar}, 
& [F_\aut,F_\ctr] & =0,
& [F_\aut,H_\ctr] & =2F_\ctr, 
\\{} 
\label{eq:bosalg4}
[H_\aut,E_\ctr] & =2E_\ctr, 
& [H_\aut,F_\ctr] & =-2F_\ctr,
& [H_\aut,H_\ctr] & =0, 
\\{} 
\label{eq:bosalg5}
[H_\ctr,E_\ctr] & =0, 
& [H_\ctr,F_\ctr] & =0, 
& [E_\ctr,F_\ctr] & =0,
\end{align}
while the coproduct relations take the following form 
\begin{align}
\label{eq:boscop1}
\copro E_\aut & =E_\aut\otimes1+q^{- H_\ctr} \otimes E_\aut
-\hbar(H_\aut+ \xi  H_\ctr  )q^{-  H_\ctr }\otimes E_\ctr
,\\
\label{eq:boscop2pre}
\copro F_\aut & =F_\aut\otimes q^{ H_\ctr}
+1\otimes F_\aut+ \hbar F_\ctr\otimes q^{  H_\ctr }(H_\aut+ \xi  H_\ctr  )
,\\
\label{eq:boscop2}
\copro H_\aut & =H_\aut\otimes 1+1\otimes H_\aut
,\\
\label{eq:boscop3}
\copro E_\ctr & =E_\ctr\otimes1+q^{- H_\ctr}\otimes E_\ctr
,\\
\label{eq:boscop4pre}
\copro F_\ctr & =F_\ctr\otimes q^{ H_\ctr}+1\otimes F_\ctr
,\\
\label{eq:boscop4}
\copro H_\ctr & =H_\ctr\otimes 1+1\otimes H_\ctr 
.\end{align}
%

\paragraph{Parameters.}

One relevant point concerning the above relations 
is that the parameters $\hbar$ and $\xi$
consistently appear as prefactors of the generators $\set{E_\ctr,F_\ctr,H_\ctr}$. 
This implies that $\hbar$ can be eliminated from the algebra
and coalgebra by the rescaling
\begin{align}
(E_\ctr,F_\ctr,H_\ctr)&\to \hbar^{-1}(E_\ctr,F_\ctr,H_\ctr),
&
\xi&\to \hbar \xi.
\end{align}
Therefore $\hbar$ is not a parameter of the Hopf algebra 
but merely of the presentation given above.
Nevertheless we refrain from removing the parameter because
it will be useful for later comparisons.

Unlike $\hbar$, the parameter $\xi$ 
cannot be removed from the algebra and coalgebra relations (at the same time)
by a redefinition of generators.
Note, however, that $\xi$ can be eliminated from the algebra relations
(see \cite{Beisert:2016qei} in conjunction with \secref{sec:sl2+d})
by the redefinition
\begin{align}
E'_\aut&= E_\aut -\xi \.Y E_\ctr,
&
F_\aut'&= F_\aut-\xi \.Y F_\ctr
\end{align}
with
\begin{align}\label{eq:ytrans}
Y&\eqq
\frac{ \half\hbar \brk!{q^{H_\ctr}-q^{-H_\ctr}}H_\ctr
  -4\hbar \sqrt{X}\sqrt{1+\hbar^2 X}\operatorname{arsinh} \brk!{\hbar\sqrt{X}}}
{\brk!{q^{H_\ctr/2}-q^{-H_\ctr/2}}^2-4\hbar^2 X} -1
\nln
&\phantom{:}=\sfrac{1}{3}\hbar^2(E_\ctr F_\ctr+\half H_\ctr^2)+\Order(\hbar^4),
\end{align}
and the invariant element 
\[\label{eq:qcasimir}
 X=  E_\ctr F_\ctr + \frac{\brk*{q^{H_\ctr/2} -q^{-H_\ctr/2}}^2}{4\hbar^2}.
\]
While this redefinition removes $\xi$ from the commutator $[E_\aut,F_\aut]$,
it does not eliminate it from $\copro E_\aut$ and $\copro F_\aut$;
in fact it introduces many additional terms.
Therefore $\xi$ is a non-trivial parameter of the Hopf algebra. 

In fact, it is not surprising to find one remaining deformation parameter $\xi$:
The original algebra $\alg{sl}(2)\directsum\alg{sl}(2)$
admits two independent q-deformation parameters $\epsilon \hbar$ and $\tilde\epsilon \hbar$.
One of them is used up in the contraction limit
\unskip\footnote{The balance of parameters for contractions can be understood as follows:
The contraction requires one parameter which is taken to zero.
If the contraction parameter was a genuine parameter of the original algebra, 
it is eliminated as a parameter of the contracted algebra.
If the contraction parameter was merely a parameter of the original algebra's presentation, 
the contracted algebra has a new continuous automorphism. 
The latter can be understood as the limit of the algebra isomorphisms 
which relate the different presentations of the original algebra.}
and it merely appears as the parameter $\hbar$ of the presentation.
The other one survives in the contraction limit as $\xi$.

\medskip

The above relations reduce to those of undeformed $\env(\alg{sl}(2)\ltimes\Complex^3)$
in the further contraction limit $\hbar\to 0$, $q=e^\hbar\to 1$,
where $\alg{sl}(2)$ and $\Complex^3$ are generated
by $\set{E_\aut,F_\aut,H_\aut}$ and $\set{E_\ctr,F_\ctr,H_\ctr}$, respectively. 
In this sense the algebra can be viewed as a one-parameter deformation of the 3D Poincar\'e algebra
and we will denote it by
\[\label{eq:algdef}
\kap_{\xi}(\alg{sl}(2)\ltimes \Complex^3)
=
\kap_{\xi}(\alg{iso}(3)).
\]
%

\subsection{Universal R-matrix}

q-deformed Hopf algebras based on simple Lie algebras possess 
a quasi-triangular structure. A natural question in this context 
is whether the quasi-triangular structure 
of q-deformed $\alg{sl}(2)\directsum\alg{sl}(2)$ survives 
the above contraction limit. 
If it does, we would like to obtain its universal R-matrix.

The universal R-matrix of $\env_{\epsilon\hbar}(\alg{sl}(2))$ 
is given by \cite{Drinfeld:1985rx,Drinfeld:1986in,Jimbo:1985zk,Jimbo:1985vd}
\[
\label{eq:r-matsl2}
\rmat_{\alg{sl}(2)}=
\exp_{-2\epsilon\hbar}\brk[s]*{(q^{\epsilon}-q^{-\epsilon})\.E\otimes F}
\exp\brk!{\half \epsilon \hbar \.  H\otimes H},
\]
where the q-exponential is defined via the q-number and q-factorial
\begin{align}
\label{eq:qnumfacexp}
[n]_\hbar&\eqq \frac{1-q^n}{1-q},
&
[n]_\hbar!&\eqq \prod_{k=1}^n [k]_\hbar,
&
\exp_\hbar[X]&\eqq \sum_n \frac{X^n}{[n]_\hbar!}.
\end{align}
The R-matrix of $\env_{\epsilon\hbar}\brk*{\alg{sl}(2)}\otimes \env_{\tilde{\epsilon}\hbar}\brk*{\alg{sl}(2)}$ 
is the product of the individual R-matrices
\unskip\footnote{Another conceivable choice of R-matrix for the combined
algebra is to take the inverse opposite R-matrix for one of its factors.
However, this turns out not to lead to a finite limit.}
\begin{align}
\rmat
&=
\rmat_{\alg{sl}(2)}\cdot \tilde\rmat_{\alg{sl}(2)}
\nln
&=
\label{eq:sl2+sl2R-mat}
\exp_{-2\epsilon\hbar}\brk[s]!{(q^{\epsilon}-q^{-\epsilon})\.E\otimes F}
\cdot
\exp_{-2\tilde\epsilon\hbar}\brk[s]!{(q^{\tilde{\epsilon} }-q^{-\tilde{\epsilon}})\.\tilde{E}\otimes \tilde{F}}
\nln&\quad\cdot
\exp\brk[s]!{\half \epsilon \hbar \. H\otimes H}
\exp\brk[s]!{\half \tilde{\epsilon} \hbar \.  \tilde{H}\otimes \tilde{H}}
\nln
&=
\exp_{-2\epsilon\hbar}\brk[s]*{\frac{q^{\epsilon}-q^{-\epsilon}}{\epsilon^2}E_\ctr\otimes F_\ctr}
\nln
&\quad\cdot
\exp_{-2\tilde\epsilon\hbar}\brk[s]*{\frac{q^{\tilde{\epsilon} }-q^{-\tilde{\epsilon}}}{\epsilon^2}
\brk*{ E_\ctr- \epsilon E_\aut} \otimes \brk*{F_\ctr-\epsilon F_\aut}}
\nln&\quad\cdot
\exp\brk[s]*{\frac{\hbar}{2\epsilon} H_\ctr\otimes H_\ctr}
\cdot\exp\brk[s]*{\frac{\tilde{\epsilon}\hbar}{2\epsilon^2}( H_\ctr-\epsilon H_\aut)\otimes ( H_\ctr-\epsilon H_\aut)},
\end{align}
where we have used the transformation
\eqref{eq:sl(2)+sl(2)DefLP,-,eq:sl(2)+sl(2)DefHC}.  
Each exponential term contains divergences in the limit $\epsilon \to 0$
paired with a simultaneous removal of the deformation.
It therefore requires some work to extract the overall divergences
of the terms and determine whether they cancel in the above combination.
In order to combine the two q-exponents into a single exponential function 
we introduce the so-called q-dilogarithm \cite{Faddeev:1993rs,Kirillov:1994en}
\unskip\footnote{Note that in \cite{Kirillov:1994en} the
q-exponential and q-dilogarithm functions are defined as
$\exp_\hbar\brk[s]{(1-q)^{-1}X}$ and $\log\exp_\hbar\brk[s]{(1-q)^{-1}X}$ respectively.
The q-dilogarithm function is defined in terms of its series expansion in
definition 9 of \cite{Kirillov:1994en}, while the relation to the q-exponential
function is shown in lemma 8.}
as the (ordinary) logarithm of the q-exponent
\[
\label{eq:q-Dilog}
\log \exp_\hbar[X]
=\sum_{n=1}^{\infty}\frac{(1-q)^{n-1}}{n\.[n]_\hbar}X^{n}.
\]
We will also require the expansion of the q-dilogarithm close to $q = 1$ 
(see corollary 10 of \cite{Kirillov:1994en})
\[
\log \exp_\epsilon\brk[s]*{\frac{X}{\epsilon}}
=-\frac{1}{\epsilon}\Li_{2}\brk{-X}
+\Order(\epsilon).
\]
We can now calculate the limit of the R-matrix. 
The expansion of the logarithm of the first q-exponential
in \eqref{eq:sl2+sl2R-mat} is
\[
\log\exp_{-2\epsilon\hbar}\brk[s]*{\frac{q^{\epsilon}-q^{-\epsilon}}{\epsilon^2}E_\ctr\otimes F_\ctr}
=\frac{1}{2\epsilon \hbar}\Li_{2}\brk!{4\hbar^2 E_\ctr\otimes F_\ctr}+\Order(\epsilon),
\]
while for the second q-exponential we find
\begin{align}
&\quad\log\exp_{-2\tilde{\epsilon}\hbar}\brk[s]*{\frac{q^{\tilde{\epsilon} }-q^{-\tilde{\epsilon}}}{\epsilon^2}
\brk*{E_\ctr- \epsilon E_\aut}\otimes \brk*{F_\ctr-\epsilon F_\aut}}
\nln
&=-\brk*{\frac{1}{2 \epsilon \hbar}+\frac{\xi}{2\hbar }}
\Li_{2}\brk!{4\hbar^2 E_\ctr\otimes F_\ctr }
\nln
&\quad
-\frac{1}{2\hbar}\brk!{ 
E_\ctr\otimes F_\aut +E_\aut\otimes F_\ctr+2\xi E_\ctr\otimes F_\ctr
}\frac{\log\brk!{1 -4\hbar^2 E_\ctr\otimes F_\ctr}}{E_\ctr\otimes F_\ctr}
+\Order(\epsilon).
\end{align}
As the exponents of the two q-exponentials commute (before taking 
the $\epsilon \to 0$ limit) the logarithm of their product is simply given
by the sum of their logarithms. It immediately follows that the
divergences of the two q-exponentials cancel, leaving
a finite contribution to the R-matrix.
The limit of the exponential functions of the Cartan generators 
is straight-forward, such that the universal R-matrix 
takes the following finite form in the limit $\epsilon\to 0$:
\begin{align}
\label{eq:bosrmat}
\rmat
&=
\exp\brk[s]*{ 
-\frac{\xi}{2\hbar }\Li_{2}\brk!{4\hbar^2 E_\ctr\otimes F_\ctr} 
-\frac{\xi}{\hbar}\log\brk!{1 -4\hbar^2 E_\ctr\otimes F_\ctr}
}
\nln &\quad
\cdot\exp\brk[s]*{ -\frac{1}{2\hbar}
\brk!{ E_\ctr\otimes F_\aut +E_\aut\otimes F_\ctr}
\frac{\log\brk!{1 -4\hbar^2 E_\ctr\otimes F_\ctr}}{E_\ctr\otimes F_\ctr}
 }
\nln &\quad
\cdot\exp\brk[s]!{\half\hbar \brk{H_\ctr\otimes H_\aut + H_\aut\otimes H_\ctr+\xi H_\ctr\otimes H_\ctr}}.
\end{align}

Let us make two remarks on the form of the resulting R-matrix:
First, the R-matrix contains both ordinary dilogarithms $\Li_2(x)$
and functions $\log(1-x)/x=-\Li'_2(x)$ within its exponents.
These function may appear unusual at first sight, 
but given that a q-exponential
can be expressed as the exponential of a q-dilogarithm, see \eqref{eq:q-Dilog},
their appearance is less surprising.
Second, the resulting R-matrix is no longer factorised into 
two constituent R-matrices after taking the limit.
The reason for the loss of factorisation is that
the constituent R-matrices are divergent on their own,
but their combination remains finite in the contraction limit.
This behaviour is analogous to the factorisation behaviour of the algebra which starts out
as the direct sum $\alg{so}(4)=\alg{sl}(2)\directsum\alg{sl}(2)$
and becomes indecomposable 
in the contraction $\alg{iso}(3)=\alg{sl}(2)\ltimes\Complex^3$.

\section{Relation to kappa-Poincar\'e}
\label{sec:kappoincare}

Before discussing maximally extended $\alg{sl}(2|2)$ in \secref{sec:sl2+d},
in this section we explore the Hopf algebra and universal R-matrix of \secref{sec:sl2+sl2} in more detail,
including its algebraic structure along with some physical implications.
The one-parameter
Hopf algebra $\kap_\xi(\alg{iso}(3))$ constructed in the previous section
\eqref{eq:bosalg0,-,eq:boscop4} as a contraction of $\env_{\epsilon
\hbar}(\alg{sl}(2)) \otimes \env_{ \tilde \epsilon \hbar}(\alg{sl}(2))$ is in
fact a one-parameter deformation of the well-known 3D kappa-Poincar\'e
algebra, first considered explicitly in \cite{Giller:1992xg} as the 3D analog
of the 4D kappa-Poincar\'e algebra of
\cite{Lukierski:1991pn,Lukierski:1992dt}.

\subsection{Comparison}

To compare our one-parameter Hopf algebra with 3D kappa-Poincar\'e we start
by introducing
$\set{L_0,L_1,L_2}$ and $\set{P_0,P_1,P_2}$
as the canonical rotation and momentum generators of $\alg{iso}(3)$ 
along with the following linear combinations
\begin{align}
L_\pm &\eqq \half( L_1 \pm i L_2), 
&
P_\pm &\eqq \half( P_1 \pm i P_2).
\end{align}
The new generators are related to those of \secref{sec:sl2+sl2} as follows
\begin{align}
H_\ctr & = 2 i \Ad_\rot P_0 &
H_\aut & = 2 i \Ad_\rot L_0
\nln \label{eq:cstoinv0}
& = 2 i P_0 &
& = 2 i L_0,
\\
E_\ctr  & = 2\Ad_\rot P_+, &
E_\aut & = 2 \Ad_\rot L_+
\nln
& = 2q^{-i P_0} P_+ & 
& = 2 q^{-iP_0}\brk!{L_+ - i \hbar P_+ \brk{L_0 + \xi P_0}},
\\
F_\ctr & = 2\Ad_\rot P_- , &
F_\aut & = 2\Ad_\rot L_- 
\nln\label{eq:cstoinv1}
& = 2q^{iP_0} P_-, & 
& = 2q^{iP_0}\brk!{L_- + i \hbar P_- \brk{L_0 + \xi P_0}},
\end{align}
with the adjoint action $\Ad_T$ and generator $T$ defined as
\begin{align}
\Ad_\rot a &\eqq \rot a \rot^{-1}, 
&
\rot &\eqq q^{P_0 L_0 + \xi P_0^2/2}.
\end{align}
In this basis the non-vanishing
commutation relations \eqref{eq:bosalg0,-,eq:bosalg5} are
given by
\begin{align}\label{eq:comnew0}
\comm{L_+}{L_-} & = \frac{i}{4} (q^{2i P_0} + q^{- 2i P_0} ) \brk{L_0 + \xi P_0} 
- \frac{\xi}{8\hbar}(q^{2iP_0} - q^{-2i P_0}) , 
\\
\comm{L_0}{L_\pm} &= \mp i L_\pm , 
\\
\comm{L_\pm}{P_\mp} & = \pm \frac1{8\hbar} (q^{2iP_0} - q^{-2i P_0}), 
\\\label{eq:comnew1}
\comm{L_0}{P_\pm} &= \comm{P_0}{L_\pm} = \mp i P_\pm,
\end{align}
such that in the limit $\hbar \to 0$ we recover the algebra $\alg{iso}(3)$, with
the three rotations $L_\mu$ and translations $P_\mu$ satisfying
\begin{align}
\label{eq:3dpoin}
\comm{P_\mu}{P_\nu} &= 0, 
&
\comm{L_\mu}{L_\nu} &= \epsilon_{\mu\nu}{}^{\rho} L_\rho, 
&
\comm{L_\mu}{P_\nu} &= \epsilon_{\mu\nu}{}^{\rho} P_\rho.
\end{align}
Here we have introduced the anti-symmetric tensor $\epsilon_{\mu\nu\rho}$ with
$\epsilon^{012} = 1$ and we contract the indices $\mu,\nu,\ldots$ with
$\eta_{\mu\nu} = \diag\brk{-1,1,1}_{\mu\nu}$. 
We will denote this algebra as $\alg{iso}(3)$
despite the apparent choice of signature (which is irrelevant 
in the complexified algebra).

The two quadratic Casimirs of the 3D Poincar\'e both have generalisations in
the deformed algebra. The first is given by \eqref{eq:qcasimir}, which in
the new basis is
\begin{align}
X = 4 P_+ P_- + \frac{(q^{i P_0} - q^{-i P_0})^2}{4\hbar^2},
\end{align}
and hence generalises the classical momentum invariant.
The second invariant element takes the form
\begin{align}
\tilde X = 4 P_+ L_- + 4 P_- L_+ 
+ \frac{i}{2\hbar}(q^{2iP_0} - q^{-2iP_0}) (L_0 + \xi P_0) 
- \frac{\xi}{2\hbar^2}(q^{i P_0} - q^{- i P_0})^2.
\end{align}
The classical limit of $\tilde X$ is $2P_\mu L^\mu$. This scalar is the
3D analogue of the 4D Pauli--Luba\'nski vector,
and hence is a measure of the spin.

The adjoint action $\Ad_T$ in the redefinitions \eqref{eq:cstoinv0,-,eq:cstoinv1} does not alter
the commutation relations, however it does modify the coproduct, which in the new
basis is given by
\begin{align}\label{eq:coprodnew0}
\copro(P_0) &= P_0 \otimes 1 + 1 \otimes P_0, 
\\
\copro(L_0) &= L_0 \otimes 1 + 1 \otimes L_0,
\\
 \copro(P_\pm) &= P_\pm \otimes q^{iP_0} + q^{-iP_0} \otimes P_\pm, 
\\ 
 \copro(L_\pm) &= L_\pm \otimes q^{iP_0} + q^{-iP_0} \otimes L_\pm
\nln \label{eq:coprodnew1}
&\quad
+ i \hbar \brk[s]!{P_\pm \otimes q^{iP_0} \brk{L_0 + \xi P_0} 
- \brk{L_0 + \xi P_0} q^{-iP_0} \otimes P_\pm},
\end{align}
while the R-matrix \eqref{eq:bosrmat} takes the form
\begin{align}
\rmat & = \Ad_{\rot \otimes \rot} 
\brk[c]*{\exp\brk[s]*{
-\frac{\xi}{2\hbar} \Li_2(16\hbar^2 P_+ \otimes P_-) - \frac{\xi}{\hbar} \log(1 -16 \hbar^2 P_+ \otimes P_-)}}
\nln & \quad
\cdot \Ad_{\rot \otimes \rot}
\brk[c]*{ \exp \brk[s]*{-\frac{1}{2\hbar} (P_+ \otimes L_- + L_+ \otimes P_-)
\frac{\log(1 - 16 \hbar^2 P_+ \otimes P_-)}{P_+ \otimes P_-}}}
\nln \label{eq:rmatnew} & \quad
\cdot \exp\brk[s]!{ - 2\hbar \brk{P_0 \otimes L_0 + L_0\otimes P_0 + \xi P_0 \otimes P_0}} .
\end{align}

Setting $\xi = 0$ we find that the defining relations of
$\kap_0(\alg{iso}(3))$ are equivalent to those of the 3D kappa-Poincar\'e
algebra \cite{Giller:1992xg,Celeghini:1990xx}.
\unskip\footnote{Note that for $\xi = 0$ the contraction of \secref{sec:sl2+sl2},
that is the limit from $\env_{\epsilon \hbar} (\alg{sl}(2)) \otimes
\env_{-\epsilon \hbar}(\alg{sl}(2))$ to  $\kap_0(\alg{iso}(3))$, was first
found in \cite{Celeghini:1990xx}. There the limit was also applied to the
universal R-matrix, however we have been unable to relate the result of
\cite{Celeghini:1990xx} to \eqref{eq:bosrmat}.}
To explicitly match the canonical presentation of the latter
\cite{Lukierski:1993wxa,Maslanka} (usually given in terms of the parameter
$\kappa$) one should set $\hbar = \half \kappa^{-1}$.  It then follows that
$\kap_\xi(\alg{iso}(3))$ \eqref{eq:algdef} is a one-parameter deformation
of the 3D kappa-Poincar\'e Hopf algebra. As discussed in
\secref{sec:sl2+sl2}, in contrast to the parameter $\hbar$ or equivalently
$\kappa$, $\xi$ is a genuine parameter of the Hopf algebra, i.e.\ it
cannot be removed by a redefinition of generators \cite{Beisert:2016qei}. 

The parameter $\xi$ can however be removed from the algebra relations of
$\kap_\xi(\alg{iso}(3))$ via the transformation \eqref{eq:ytrans}.
Furthermore, it was shown in \cite{Borowiec:2009vb} that there is an analogous
transformation mapping the algebra relations of $\kap_0(\alg{iso}(3))$ to those
of the undeformed 3D Poincar\'e algebra. It therefore follows that the
algebra relations of $\kap_\xi(\alg{iso}(3))$ can as well be mapped to
those of $\env(\alg{iso}(3))$. It is important to note that this does not give
a map between the Hopf algebras $\kap_\xi(\alg{iso}(3))$ and
$\env(\alg{iso}(3))$ as the transformation will generate many additional terms
in the coproduct depending on $\xi$ and $\hbar$.

In this paper we are working with algebras over the complex numbers,
however discussions of kappa-Poincar\'e and kappa-Euclidean algebras
typically focus on particular real forms.
Therefore let us briefly comment on the possible
real forms \cite{Twietmeyer:1991mj} of the deformed 3D kappa-Poincar\'e
algebra. A number of the most common real forms can
be extended to include the new parameter $\xi$ upon imposing a suitable
reality condition. We have checked examples of mixed and definite signature,
both with $q \in \Real$ and $|q| = 1$.  In all cases we find that either
$\xi \in \Real$ or $\xi \in i \Real$ with the former descending
from corresponding real forms of $\env_{\epsilon\hbar}(\alg{sl}(2)) \otimes
\env_{\tilde \epsilon \hbar}(\alg{sl}(2))$, and the latter only appearing after
taking the limit.

\subsection{Classical limit}

In \secref{sec:sl2+sl2} the parameter $\xi$ came from an asymmetry \eqref{eq:asymlimit} 
in the contraction limit of $\env_{\epsilon \hbar}(\alg{sl}(2)) \otimes
\env_{\tilde \epsilon \hbar}(\alg{sl}(2))$. We can further
clarify the role of $\xi$ by considering the classical limit of the Hopf
algebra. Let us introduce the standard expressions for the cobracket $\cobrack$
and classical r-matrix $r$
\begin{align}\label{eq:coprmatex}
\copro(a) - \copro^{\cop}(a) &\eqq 2 \hbar \delta(a) + \Order(\hbar^2),
&
\rmat &\eqq 1 \otimes 1 + 2\hbar r + \Order(\hbar^2),
\end{align}
such that the classical r-matrix generates the cobracket through the coboundary
condition
\begin{equation}\label{eq:coboundary}
\comm{a \otimes 1 + 1 \otimes a}{r} = \delta(a).
\end{equation}
Introducing the anti-symmetrised and symmetrised tensor products, $a \wedge b =
a \otimes b - b \otimes a$ and $a \odot b = a \otimes b + b \otimes a$, we
expand the coproduct \eqref{eq:coprodnew0,-,eq:coprodnew1} to first order in $\hbar$ to find
the following cobracket
\begin{align}\label{eq:cobracket0}
\cobrack(P_0) &= \cobrack(L_0) = 0, 
\\
\cobrack(P_\pm) &= i P_\pm \wedge P_0, 
\\\label{eq:cobracket1}
\cobrack(L_\pm) &= i L_\pm \wedge P_0 + i P_\pm \wedge \brk{L_0 + \xi P_0}.
\end{align}
Similarly expanding \eqref{eq:rmatnew} the classical r-matrix takes the form
\begin{equation}\label{eq:classrmat}
r = 2\brk{ P_+ \wedge L_- - P_- \wedge L_+ + \xi P_+ \wedge P_- }
+  \brk { P_\mu \odot L^\mu 
+ \half \xi P_\mu \odot P^\mu }.
\end{equation}
One can check explicitly that the coboundary condition
\eqref{eq:coboundary} is satisfied and further that the classical r-matrix
\eqref{eq:classrmat} solves the classical Yang--Baxter equation
\begin{equation}\label{eq:cybe}
[[r,r]] \eqq [r_{12},r_{13}] + [r_{12},r_{23}] + [r_{13},r_{23}] = 0.
\end{equation}

The classical r-matrix \eqref{eq:classrmat} takes a form that resembles
the Drinfel'd--Jimbo solution for a simple Lie algebra
\cite{Drinfeld:1985rx,Drinfeld:1986in,Jimbo:1985zk,Jimbo:1985vd}.
The symmetric part,
\begin{equation}\label{eq:casimir}
\Casimir = 
P_\mu \odot L^\mu 
+ \half \xi P_\mu \odot P^\mu,
\end{equation}
is an $\alg{iso}(3)$-invariant element of $\alg{iso}(3)^{\odot 2}$, i.e.\ a
quadratic Casimir.  In contrast to the situation for simple Lie algebras, the
space of quadratic Casimirs of $\alg{iso}(3)$ is two dimensional.  This follows
from the fact that $\alg{iso}(3)$ is a contraction of the direct sum algebra
$\alg{so}(4) = \alg{sl}(2) \directsum \alg{sl}(2)$, which by definition has a
two-dimensional space of quadratic Casimirs.  A basis of quadratic Casimirs of
$\alg{iso}(3)$ is given by $P_\mu \odot L^\mu$ and $P_\mu \odot P^\mu$, and
hence $\hbar$ and $\xi$ parameterise an arbitrary
element of this space.

Substituting the Casimir \eqref{eq:casimir} into the left-hand side
of the classical Yang--Baxter equation \eqref{eq:cybe} we find
\begin{align}\label{eq:omegadef}
[[\Casimir,\Casimir]] &= - \omega, 
&
\omega &= -
\half \epsilon^{\mu\nu\rho}(P_\mu \wedge P_\nu \wedge L_\rho + \sfrac{2}{3}
\xi P_\mu \wedge P_\nu \wedge P_\rho),
\end{align}
where the $\Order (\xi^2)$ term vanishes as the $P_\mu$ commute amongst themselves.
By definition $\omega$ is an $\alg{iso}(3)$-invariant element of $\alg{iso}(3)^{\wedge 3}$.
This is also a two-dimensional space \cite{Stachura}, with
$\epsilon^{\mu\nu\rho} P_\mu \wedge P_\nu \wedge L_\rho$ and
$\epsilon^{\mu\nu\rho} P_\mu \wedge P_\nu \wedge P_\rho$ forming a basis,
such that $\hbar$ and $\xi$ again parameterise an arbitrary element.

It is now the anti-symmetric part of the classical r-matrix
\eqref{eq:classrmat} that generates the cobracket in \eqref{eq:coboundary}, and
hence generates the deformation of the algebra.  This is given by
\begin{equation}\label{eq:classrmatdj} \hat r = 2 (P_+ \wedge L_-
- P_- \wedge L_+ + \xi P_+ \wedge P_- ), \end{equation}
which solves the modified classical Yang--Baxter equation
\begin{equation}\label{eq:mcybe}
[[\hat r, \hat r]] = \omega, 
\end{equation}
such that $\hat r + \Casimir$ solves the classical Yang--Baxter equation.

It follows from \eqref{eq:casimir,-,eq:mcybe} that the term $2\xi P_+ \wedge
P_-$ in the classical r-matrix \eqref{eq:classrmat} does not correspond to a
Drinfel'd twist of the standard 3D kappa-Poincar\'e algebra.  Indeed this
would imply that $r - 2\xi P_+ \wedge P_-$ also satisfies the classical
Yang--Baxter equation. One can easily see this is not the case as there will no
longer be a term on the right-hand side of \eqref{eq:mcybe} linear in $\xi$
cancelling the corresponding term in \eqref{eq:omegadef}.

\subsection{Higher-dimensional kappa-Poincar\'e algebras}

In the analysis of the classical limit we have seen that the 3D Poincar\'e algebra has certain
special algebraic features.  In order to understand the importance of these, we
now consider to what extent the considerations above can be extended to the
kappa-Poincar\'e algebra in arbitrary dimension
\cite{Lukierski:1993wxa,Maslanka}.  As the $d$-dimensional Poincar\'e algebra
$\alg{iso}(d)$ can be found as a contraction of $\alg{so}(d+1)$, the
$d$-dimensional kappa-Poincar\'e algebra should be found as an analogous
contraction of $\env_\hbar(\alg{so}(d+1))$. As we have seen this is indeed the
case for $d = 3$. It has also been shown explicitly for $d = 2$
\cite{Celeghini:1990bf} and $d = 4$ \cite{Lukierski:1992dt}. However, in these
two cases the limit leads to divergences in the universal R-matrix.

The finite limit of the R-matrix for $d = 3$ has its origin in the
non-simplicity of $\alg{so}(4)$. In consequence, the R-matrix of 
$\env_\hbar(\alg{so}(d+1))$ factorises for $d=3$ into commuting 
$\alg{sl}(2)$ R-matrices
$\rmat_{\alg{so}(4)}=\rmat_{\alg{sl}(2)}\tilde{\rmat}_{\alg{sl}(2)}$.  
Given an R-matrix $\mathcal{R}$ of a quasi-triangular Hopf algebra, 
taking the inverse and transpose $(\rmat^{-1})^{\cop} $ also gives 
an R-matrix of that Hopf algebra. Therefore, for 
$\env_\hbar(\alg{so}(4)) = \env_\hbar(\alg{sl}(2))\otimes\env_\hbar(\alg{sl}(2))$ 
we have the two R-matrices 
\begin{align}
\rmat_{\alg{so}(4)}&=\rmat_{\alg{sl}(2)}\tilde \rmat_{\alg{sl}(2)},
&
(\rmat_{\alg{so}(4)}^{-1})^\cop&=(\rmat_{\alg{sl}(2)}^{-1})^{\cop} (\tilde{\rmat}_{\alg{sl}(2)}^{-1})^\cop,
\end{align}
but in addition, since both $\alg{sl}(2)$ parts are independent, 
there are also another pair of R-matrices of $\env_\hbar(\alg{so}(4))$
\begin{align}
\rmat'_{\alg{so}(4)}&=\rmat_{\alg{sl}(2)}(\tilde{\rmat}_{\alg{sl}(2)}^{-1})^\cop ,
&
(\rmat'^{-1}_{\alg{so}(4)})^\cop&=(\rmat_{\alg{sl}(2)}^{-1})^{\cop}\tilde{\rmat}_{\alg{sl}(2)}.
\end{align}
The latter two have a finite contraction limit, while the former two diverge.
Indeed the R-matrix \eqref{eq:sl2+sl2R-mat} with $\xi = 0$ is
of the latter type. To see this let us consider $\xi = 0$, in
which case the starting point of \secref{sec:sl2+sl2} is the Hopf algebra
$\env_{\epsilon\hbar}(\alg{sl}(2))\otimes \env_{-\epsilon\hbar}(\alg{sl}(2))$
with an R-matrix of the type 
$\rmat_{\alg{sl}(2)}(\epsilon\hbar) \tilde \rmat_{\alg{sl}(2)}(-\epsilon\hbar)$ 
where we now indicate the dependence of the R-matrices on the deformation parameter.
The Hopf algebras $\env_{-\hbar}(\alg{sl}(2))$ and $\env_{\hbar}(\alg{sl}(2))$ 
are isomorphic, where the isomorphism, however, maps $\rmat(-\hbar) \to (\rmat^{-1}(\hbar))^{\cop}$.
Thus re-expressing the R-matrix \eqref{eq:sl2+sl2R-mat} on
$\env_{\epsilon \hbar}(\alg{sl}(2))\otimes \env_{+\epsilon\hbar}(\alg{sl}(2))$
we indeed find that it is of the type $\rmat_{\alg{sl}(2)}(\tilde{\rmat}_{\alg{sl}(2)}^{-1})^\cop$.

\bigskip

In order to clarify these results let us take $d\neq 3$ and assume we have a contraction of the
Hopf algebra $\env_{\hbar}(\alg{so}(d+1))$ to $\kap(\alg{iso}(d))$.
We further assume that there is a classical limit, $\hbar \to 0$, in which
we have a contraction of $\alg{so}(d+1)$ to $\alg{iso}(d)$, and the
cobracket generating the deformation $\env_{\hbar}(\alg{so}(d+1))$ contracts to
that generating $\kap(\alg{iso}(d))$.
Splitting the generators of $\alg{so}(d+1)$ into those of an $\alg{so}(d)$
subalgebra, $\tilde L_{\mu \nu} = \tilde L_{[\mu\nu]}$, $\mu, \nu =0,\ldots d-1$ and the rest,
$P_\mu$, the contraction to $\alg{iso}(d)$ is given by rescaling $P_\mu
\to \epsilon^{-1} P_\mu$ and taking $\epsilon \to 0$.
In the limit $\tilde L_{\mu\nu}$ are then the rotations and $P_\mu$ the
translations of the $d$-dimensional Poincar\'e algebra.
The cobracket of kappa-Poincar\'e takes the form
\cite{Lukierski:1993wxa,Maslanka}
\begin{align}
\label{eq:cobracketd}
\delta(P_\mu) &= n^\nu P_\mu \wedge P_\nu, 
&
\delta(\tilde L_{\mu\nu}) &= - n_{\mu}\tilde L_{\nu \rho} \wedge P^\rho + n_{\nu}\tilde L_{\mu \rho} \wedge P^\rho,
\end{align}
where $n^\mu$ is a fixed vector.  Recalling that the cobracket comes with a
power of $\hbar$ in the expansion of the coproduct \eqref{eq:coprmatex}, the
expressions \eqref{eq:cobracketd} imply that, in addition to
rescaling $P_\mu \to \epsilon^{-1} P_\mu$, we should also rescale $\hbar \to
\epsilon \hbar$ for the contraction to be well-defined.

For $d \neq 3$ the algebra $\alg{so}(d+1)$ is simple and has a single quadratic
Casimir.  The leading term of this Casimir in the contraction is the
quadratic Casimir of $\alg{iso}(d)$
\[\label{eq:casimird}\Casimir_d = P_\mu \odot P^\mu.\] 
Since the symmetric part of the Drinfel'd--Jimbo classical r-matrix for a
simple Lie algebra is the quadratic Casimir, its leading term in the
contraction limit will contain $\Casimir_d$ and hence is  quadratic in $P_\mu$.
However, as the classical r-matrix comes with a power of $\hbar$ in the
expansion of the R-matrix \eqref{eq:coprmatex}, for a finite limit it should be
at most linear in $P_\mu$. Therefore, the classical r-matrix necessarily
diverges in the contraction. Returning to the deformed Hopf algebras it follows
that taking the contraction limit in the Drinfel'd--Jimbo universal R-matrix
for $\env_\hbar(\alg{so}(d+1))$ is problematic for $d \neq 3$.

\bigskip

Taking a different perspective we may instead start from the solution of the modified
classical Yang--Baxter equation \eqref{eq:mcybe} that generates the
kappa-Poincar\'e deformation \cite{Lukierski:1993wxa,Maslanka,Zakrzewski}
\unskip\footnote{In 3D we have $\tilde L^{\mu\nu} =
\epsilon^{\mu\nu\rho} L_\rho$ such that \eqref{eq:rmatd} matches
\eqref{eq:classrmatdj} and \eqref{eq:omegadef} if we take $n_0 = -i$, $n_1 =
n_2 = 0$ and $\xi = 0$.}
\begin{align}\label{eq:rmatd}
\hat r_d &= n_\mu  P_\nu \wedge \tilde L^{\mu\nu}, 
&
 [[\hat r_d, \hat r_d]] &= \omega_d = -\half n^2 P_\mu \wedge P_\nu \wedge \tilde L^{\mu\nu},
\end{align}
where $\omega_d$ is an $\alg{iso}(d)$-invariant element of
$\alg{iso}(d)^{\wedge 3}$. That is the cobracket for the kappa-Poincar\'e
algebra \eqref{eq:cobracketd} obeys the coboundary condition
\eqref{eq:coboundary} with this classical r-matrix. However, it has been shown
for $d \neq 3$ that there exists no symmetric term whose sum with $\hat{r}_d$
\eqref{eq:rmatd} solves the classical Yang--Baxter equation
\cite{Lukierski:1993wxa,Maslanka}.

\bigskip

Finally returning to $d = 3$ the above analysis further clarifies that the
finite limit of the universal R-matrix is tied to the existence of a second
quadratic Casimir that is linear in $P_\mu$, which in turn is a consequence of
$\alg{iso}(3)$ being a contraction of the direct sum algebra $\alg{so}(4) =
\alg{sl}(2) \directsum \alg{sl}(2)$. Indeed, our choice of initial R-matrix
when contracting the Hopf algebra $\env_{\epsilon \hbar}(\alg{sl}(2)) \otimes
\env_{\tilde \epsilon \hbar}(\alg{sl}(2))$ should be such that
the leading term in the limit of the symmetrised classical
r-matrix is the Casimir linear in $P_\mu$.

\subsection{R-matrix and 3D scattering problem}

Having studied the algebraic structure of the universal R-matrix for
kappa-Poincar\'e symmetry, we conclude this section by asking what purpose it
may serve in a physical context.  One idea, based on integrable models in 2D,
is that the R-matrix describes a two-particle scattering process.  Let us
therefore discuss some of its implications.

We set up a state $\ket{p}\otimes \ket{q}$ with a pair of 
well-defined momenta $(p_\mu,q_\mu)$ to describe the two particles
and let the R-matrix $\rmat$ act on it. We will only be interested in 
the momenta of the particles after the scattering process.
To this end we note that the particle momenta are measured 
as the eigenvalues of the momentum generators $P_\mu$.
We therefore compute how the equivalent set of generators 
$\set{E_\ctr,F_\ctr,H_\ctr}$ commutes past the R-matrix
%
%
\begin{align}
\label{eq:rmatmom1}
\rmat^{-1} (E_\ctr\otimes 1) \rmat &= E_\ctr\otimes q^{-H_\ctr},
\\
\label{eq:rmatmom2}
\rmat^{-1} (1\otimes F_\ctr) \rmat &= q^{H_\ctr}\otimes F_\ctr,
\\
\rmat^{-1} (H_\ctr\otimes 1) \rmat &= 
H_\ctr\otimes 1 - \hbar^{-1}\log\brk!{1 -4\hbar^2 q^{H_\ctr} E_\ctr \otimes q^{-H_\ctr}F_\ctr},
\\
\rmat^{-1} (1\otimes H_\ctr) \rmat &= 
1\otimes H_\ctr + \hbar^{-1}\log\brk!{1 -4\hbar^2 q^{H_\ctr} E_\ctr \otimes q^{-H_\ctr}F_\ctr},
\\
\rmat^{-1}(F_\ctr\otimes 1) \rmat &=
F_\ctr \otimes q^{H_\ctr}
+ 1\otimes F_\ctr
- \frac{q^{ 2H_\ctr}\otimes F_\ctr}{1 -4\hbar^2 q^{H_\ctr}E_\ctr\otimes q^{-H_\ctr} F_\ctr},
\\
\rmat^{-1}(1\otimes E_\ctr) \rmat &=
q^{-H_\ctr}  \otimes E_\ctr
+ E_\ctr\otimes 1
- \frac{E_\ctr \otimes q^{-2H_\ctr}}{1 -4\hbar^2 q^{H_\ctr}E_\ctr\otimes q^{-H_\ctr} F_\ctr}.
\end{align}
The operators on the right-hand side measure the momenta
of the outgoing particles, which is thus completely fixed
in terms of the ingoing momenta on the left-hand side.
As a result, the outgoing state, curiously, has well-defined momenta
\[
\rmat \. \ket{p}\otimes \ket{q} \sim \ket{p'}\otimes \ket{q'}.
\]
This result is in contrast to the intuition that a scattering process 
in 3D (or any other number of dimensions above 2) produces 
a linear combination of states with continuously varying momenta.

For concreteness, let us express the relationship between
the ingoing momenta $(p_\mu,q_\mu)$ and outgoing momenta $(p'_\mu,q'_\mu)$ 
in the common basis $P_\mu$ of kappa-Poincar\'e symmetry. We find
\begin{align}
\label{eq:inoutmom1}
p'_+ &= e^{-iq_0/\kappa}r^{-1/2} p_+,
\\
q'_- &= e^{ip_0/\kappa}r^{-1/2} q_-,
\\
p'_0 &= p_0 +i\kappa\log r,
\\
q'_0 &= q_0 -i\kappa\log r,
\\
p'_- &=
 e^{iq_0/\kappa} r^{1/2} p_- 
+ e^{i(q_0-p_0)/2\kappa} r^{1/2} q_-
- e^{i(3p_0+q_0)/2\kappa} r^{-1/2} q_-,
\\
q'_+ &=
e^{-ip_0/\kappa} r^{1/2} q_+
+ e^{i(q_0-p_0)/2\kappa} r^{1/2} p_+ 
- e^{-i(p_0+3q_0)/2\kappa} r^{-1/2} p_+,
\label{eq:inoutmom6}
\end{align}
with
\[
r\eqq 1-\frac{4}{\kappa^2}e^{i(p_0-q_0)/2\kappa} p_+ q_-.
\]
In fact, five of these six relations are implied by conservation laws, 
namely the conservation of overall momentum
alias quasi-cocommutativity for $\set{E_\ctr, F_\ctr, H_\ctr}$
\begin{align}
\label{eq:momcons1}
p'_0 + q'_0 &= p_0 + q_0,
\\
e^{-iq'_0/2\kappa} p'_+ + e^{ip'_0/2\kappa} q'_+ &=
e^{iq_0/2\kappa} p_+ + e^{-ip_0/2\kappa} q_+ ,
\\
e^{-iq'_0/2\kappa} p'_- + e^{ip'_0/2\kappa} q'_- &=
e^{iq_0/2\kappa} p_- + e^{-ip_0/2\kappa} q_-,
\label{eq:momcons3}
\end{align}
as well as conservation of the mass shell for each particle
due the centrality of the element $X$ in \eqref{eq:qcasimir}
\begin{align}
\label{eq:massshell1}
p'_+ p'_- - \kappa^2 \sin^2\brk{p'_0/2\kappa}
&= p_+ p_- - \kappa^2 \sin^2\brk{p_0/2\kappa},
\\
q'_+ q'_- - \kappa^2 \sin^2\brk{q'_0/2\kappa}
&= q_+ q_- - \kappa^2 \sin^2\brk{q_0/2\kappa}.
\label{eq:massshell2}
\end{align}
These five relationships constrain the six outgoing momenta 
to a one-parameter family.
Nevertheless, there is a sixth relationship, which can 
be expressed in a somewhat symmetric form as
\unskip\footnote{In fact, this relationship leaves a few other discrete choices
but we chose it due to its symmetric form.
Alternatively, any one of the relationships \eqref{eq:inoutmom1,-,eq:inoutmom6}
could be used instead.}
\[\label{eq:sixthconservation}
e^{i (q'_0-p'_0)/2\kappa} p'_+ q'_-
=
e^{i (p_0-q_0 )/2\kappa} p_+  q_-.
\]
It corresponds to the combination of \eqref{eq:rmatmom1} and \eqref{eq:rmatmom2}
\[
\rmat^{-1} (E_\ctr\otimes F_\ctr) \rmat = q^{H_\ctr} E_\ctr\otimes q^{-H_\ctr} F_\ctr;
\]
in other words it follows from explicit commutation with the R-matrix.
The physical origin of this final relationship, for example,
how it follows from a hypothetical sixth conserved quantity, and
the deeper meaning of the above transformation \eqref{eq:inoutmom1,-,eq:inoutmom6}
of momenta and whether it can serve within a reasonable particle scattering process,
remain to be understood.

Note that a superficially similar scattering process has been discussed 
in the context of the AdS/CFT correspondence, 
see appendix B of \cite{Beisert:2008tw}.
More concretely, this is a 2D scattering process with 
the 2D momenta embedded into a 3D momentum vector
whose coproduct is equivalent to the one of $\set{E_\ctr,F_\ctr,H_\ctr}$.
Consequently, all of the above five conservation laws 
\eqref{eq:momcons1,-,eq:massshell2} are respected by 
this scattering problem, 
but the remaining sixth relationship \eqref{eq:sixthconservation} is manifestly different.
In our basis, it can be expressed as $p'_0-q'_0=p_0-q_0$
implying that the energies of the individual particles
are preserved across the scattering.
This implies a different transformation for the momenta
\eqref{eq:inoutmom1,-,eq:inoutmom6}.

\section{Maximally extended \texorpdfstring{$\alg{sl}(2|2)$}{sl(2|2)}
 from \texorpdfstring{$\alg{d}(2,1;\epsilon) \directsum \alg{sl}(2)$}{d(2,1;epsilon)+sl2}}
\label{sec:sl2+d}

We now turn to our primary interest, recovering the quasi-triangular Hopf
algebra of \cite{Beisert:2016qei} as a contraction limit of
$\env_\hbar(\alg{d}(2,1;\epsilon)) \otimes \env_{\tilde
\epsilon\hbar}(\alg{sl}(2))$.  
The maximally extended $\alg{sl}(2|2)$ Hopf algebra of \cite{Beisert:2016qei}
was constructed as the smallest
quasi-triangular Hopf algebra containing the centrally extended $\env_\hbar
(\alg{sl}(2|2)\ltimes \Complex^2)=\env_\hbar(\alg{psl}(2|2)\ltimes \Complex^3)$ as
a Hopf subalgebra.  The structure of this algebra has the form
$\env_{\hbar,\xi}(\alg{sl}(2)\ltimes \alg{psl}(2|2) \ltimes \Complex^3)$,
where the $\alg{sl}(2)$ factor plays the role of a continuous 
outer automorphism for the remainder of the algebra.  
As discussed at the beginning of \secref{sec:sl2+sl2} this Hopf algebra
possesses a number of unusual features, including the appearance of plain
$\hbar$ factors which are not within exponents $q=e^\hbar$, the existence of an additional free parameter $\xi$, as
well as the non-factorisable form of the universal R-matrix, which involves
logarithms and dilogarithms. In \secref{sec:sl2+sl2} we saw that for the
Hopf subalgebra $\env_{\hbar,\xi}(\alg{sl}(2) \ltimes \Complex^3)$ these
features can be understood by considering a certain contraction of
$\env_{\epsilon \hbar}(\alg{sl}(2)) \otimes \env_{\tilde
\epsilon\hbar}(\alg{sl}(2))$.  Therefore, in this section our aim is to recover the
full maximally extended $\alg{sl}(2|2)$ Hopf algebra in a similar limit.
Our starting point for this will be to promote one
$\env_{\epsilon \hbar}(\alg{sl}(2))$ factor in the construction of
\secref{sec:sl2+sl2} to $\env_\hbar(\alg{d}(2,1;\epsilon))$
while keeping the other factor as $\env_{\tilde\epsilon \hbar}(\alg{sl}(2))$.

\subsection{Lie superalgebra \texorpdfstring{$\alg{d}(2,1;\epsilon)$}{d(2,1;epsilon)}}

Let us begin by introducing the exceptional Lie superalgebra $\alg{d}(2,1;\epsilon)$.
This superalgebra depends on the continuous parameter
$\epsilon$ and hence forms a one-parameter family of Lie superalgebras.
The even subalgebra consists of three mutually commuting $\alg{sl}(2)$ algebras, i.e.\ 
$\alg{sl}(2)\directsum \alg{sl}(2)\directsum \alg{sl}(2)$. The odd 
part is spanned by 8 odd generators transforming in the tri-fundamental representation
of the even subalgebra. 
The anti-commutators of two odd generators $Q$
take the schematic form
\[
\acomm{Q}{Q}  \sim s_1 T_1+s_2 T_2 +s_3 T_3,
\]
where the $T_i$ denote normalised generators of the three $\alg{sl}(2)$
algebras. The parameters $s_i$ are constrained by the Jacobi identity to
satisfy $s_{1}+s_{2}+s_{3}=0$.


The superalgebra with parameters $s_i$ and the one with parameters $\lambda s_i$
($0\neq\lambda\in \Complex$) are isomorphic.  The isomorphism is simply given
by scaling the odd generators by $\sqrt{\lambda}\in\Complex$.  We can
therefore always normalise one parameter and write the parameters $s_i$ in
terms of a single parameter $\epsilon$. There are multiple ways to do this and our choice is
\begin{align}
\label{eq:s123param}
s_{1}&=1,
&
s_{2}&=\epsilon,
&
s_{3}&=-1-\epsilon.
\end{align}
In general we will work with the parameters $s_i$ as they preserve the symmetry
among the three $\alg{sl}(2)$ algebras.  We will however need to introduce the
parameter $\epsilon$ in order to take the contraction limit in \secref{ssec:limit}.


Considering the superalgebra $\alg{d}(2,1;\epsilon)$ by itself there are two ways of
taking the $\epsilon \to 0$ limit that will be important in our construction.
Using the parameterisation \eqref{eq:s123param} the first is to directly take
$\epsilon \to 0$ leading to $\alg{sl}(2)\ltimes\alg{psl}(2|2)$, that is $\alg{psl}(2|2)$
together with its $\alg{sl}(2)$ outer automorphism. The second involves first
rescaling $T^{(2)} \to \epsilon^{-1} T^{(2)}$ and then taking $\epsilon \to 0$.
This leads to $\alg{psl}(2|2)\ltimes\Complex^{3}$, that is the triple central
extension of $\alg{psl}(2|2)$. 

In order to combine these two limits we will introduce an additional
$\alg{sl}(2)$ algebra in the spirit of \secref{sec:sl2+sl2} to obtain 
in the limit $\alg{sl}(2)\ltimes \alg{psl}(2|2) \ltimes \mathbb{C}^3$, 
the maximally extended $\alg{sl}(2|2)$ algebra.  
Lifting this limit to the q-deformed
algebras our aim is then to find the Hopf algebra
$\env_{\hbar,\xi}(\alg{sl}(2) \ltimes \alg{psl}(2|2) \ltimes \Complex^3)$ of
\cite{Beisert:2016qei} as a contraction of $\env_{\hbar}(\alg{d}(2,1;\epsilon))\otimes
\env_{\tilde{\epsilon}\hbar}(\alg{sl}(2))$. 

\subsection{q-deformation of \texorpdfstring{$\alg{d}(2,1;\epsilon)$}{d(2,1;epsilon)}}

Let us now define the q-deformed Hopf algebra $\env_\hbar(\alg{d}(2,1;\epsilon))$
and its universal R-matrix \cite{Thys:2001}. 
The three even $\alg{sl}(2)$ subalgebras are deformed with $q^{s_i}=e^{\hbar s_i}$.
\unskip\footnote{A rescaling of the parameters $s_i$ requires an inverse 
rescaling of $\hbar$ in addition to the appropriate rescaling 
to the odd generators for the corresponding Hopf algebras to be isomorphic. 
In this sense the $\hbar$ in $\env_\hbar(\alg{d}(2,1;\epsilon))$ 
corresponds to the choice \eqref{eq:s123param}.
}
The even subalgebra is therefore given by 
\[ 
\env_{\hbar s_1}(\alg{sl}(2))\otimes
\env_{\hbar s_2}(\alg{sl}(2))\otimes
\env_{\hbar s_3}(\alg{sl}(2)).
\]
Note that their coproduct can however contain in addition to the standard coproduct 
a tail involving the odd generators.
Note further that the deformation of the $\alg{sl}(2)$
corresponding to $s_2$ will vanish in the limit $\epsilon \to 0$ 
while the deformation for the other two $\alg{sl}(2)$ subalgebras will remain.
The former $\env_{\hbar s_2}(\alg{sl}(2))$ subalgebra will thus 
replace the first $\env_{\epsilon \hbar}(\alg{sl}(2))$ Hopf algebra 
of \secref{sec:sl2+sl2}.
The latter two subalgebras $\env_{\hbar s_1}(\alg{sl}(2))$ and 
$\env_{\hbar s_3}(\alg{sl}(2))$ will become the two $\alg{sl}(2)$ subalgebras 
of $\alg{psl}(2|2)$ in the limit $\epsilon \to 0$.

\paragraph{Algebra and coalgebra.}

We define the q-deformed Hopf algebra $\env_\hbar(\alg{d}(2,1;\epsilon))$
in terms of three sets of simple generators $E_i$, $F_i$ and their 
corresponding Cartan generators $H_i$, $i=1,2,3$. The generators 
$E_2$ and $F_2$ are odd while $E_{1,3}$, $F_{1,3}$ are even. 
We will make use of the graded q-commutator
\[
\comm{a}{b}_{\alpha}\eqq ab-(-1)^{|a||b|}e^{\alpha} ba,
\]
where the degree is $|a| = 0$ for even generators and $|a| = 1$ for
odd generators. For undeformed commutators we simply write 
$\comm{a}{b}\eqq\comm{a}{b}_0$. 
The commutation relations of the simple generators are given by
\begin{align}
[H_{i},E_{j}] & =a_{ij}E_{j},
&
[H_{i},F_{j}] & =-a_{ij}F_{j}, 
&
[E_{i},F_{j}] & =\delta_{ij}\frac{q_i^{H_{i}}-q_i^{-H_{i}}}{q_i-q_i^{-1}},
\end{align}
where the Cartan matrix $a_{ij}$ and the q-exponents $d_i$ are given by
\begin{align}
a_{ij} & =
\begin{pmatrix}2 & -1 & 0\\
s_{1} & 0 & s_{3}\\
0 & -1 & 2
\end{pmatrix}_{ij} , 
& d_{i} & =\begin{pmatrix}s_{1} & -1 & s_{3}\end{pmatrix}_i.
\end{align}
The latter are chosen such that they symmetrise the Cartan matrix $d_i a_{ij}=d_j a_{ji}$. 
Furthermore, they give rise to the deformation strength of the respective simple generators
\[
q_i\eqq e^{d_i \hbar}.
\]
To define the non-simple generators and Serre relations
we introduce the left and right adjoint action 
\begin{align}
a\triangleright b &\eqq (-1)^{|b||a_{(2)}|} \.a_{(1)}\.b\. \antipode(a_{(2)}),
\\
b\triangleleft a &\eqq (-1)^{|a_{(1)}||b|} \.\antipode(a_{(1)})\.b\. a_{(2)},
\end{align}
where we made use of Sweedler's notation for the coproduct
$\copro(a)=a_{(1)} \otimes a_{(2)}$,
with an implicit sum over all terms. 
$\antipode$ denotes the antipode.

\medskip

We define the six odd non-simple generators
\begin{align}\label{eq:fermns1}
E_{12} & \eqq E_{1}\triangleright E_{2}=[E_{1},E_{2}]_{\hbar s_{1}}, 
& F_{21} & \eqq F_{2}\triangleleft F_{1}=[F_{2},F_{1}]_{-\hbar s_{1}},\\
E_{32} & \eqq E_{3}\triangleright E_{2}=[E_{3},E_{2}]_{\hbar s_{3}}, 
& F_{23} & \eqq F_{2}\triangleleft F_{3}=[F_{2},F_{3}]_{-\hbar s_{3}},\\
\label{eq:fermns2}
E_{132} & \eqq (E_{1}E_{3})\triangleright E_{2}
=[E_{1},E_{32}]_{\hbar s_{1}},
& F_{213} & \eqq F_{2}\triangleleft (F_{1}F_{3})
=[F_{23},F_{1}]_{-\hbar s_{1}}.
\end{align}
The Serre relations are then given by
\begin{align}
\label{eq:serre1}
E_2^2 & = 0, & F_2^2 & = 0,\\
E_{1}\triangleright E_{3} & =[E_{1},E_{3}]  =0, 
& F_{1}\triangleleft F_{3} & =[F_{1},F_{3}]  =0,
\\
E_{1}\triangleright(E_{1}\triangleright E_{2})
& =[E_{1},E_{12}]_{-\hbar s_{1}} =0, 
& (F_{2}\triangleleft F_{1})\triangleleft F_{1}
& =[F_{21},F_{1}]_{\hbar s_{1}}  =0,
\\
E_{3}\triangleright(E_{3}\triangleright E_{2})
& =[E_{3},E_{32}]_{-\hbar s_{3}} =0, 
& (F_{2}\triangleleft F_{3})\triangleleft F_{3}
& =[F_{23},F_{3}]_{\hbar s_{3}}  =0.
\label{eq:serre2}
\end{align}
Note that by the q-Jacobi identity the non-simple generators $E_{132}$ and $F_{213}$ 
satisfy the identities
$[E_{3},E_{12}]_{\hbar s_{3}} = [E_{1},E_{32}]_{\hbar s_1}$
and $[F_{21},F_{3}]_{-\hbar s_{3}} = [F_{23},F_{1}]_{-\hbar s_{1}}$ respectively.

\medskip

The q-deformed coalgebra is defined on the simple generators via the coproduct
\begin{align}
\copro E_{i} & =E_{i}\otimes1+q_i^{-H_{i}}\otimes E_{i},\\
\copro F_{i} & =F_{i}\otimes q_i^{H_{i}}+1\otimes F_{i},\\
\copro H_{i} & =H_{i}\otimes1+1\otimes H_{i},
\end{align}
where the tensor product is graded in the usual way
\[
(a\otimes b) (c\otimes d)=(-1)^{|b||c|}ac \otimes bd.
\]
%

\paragraph{Third \texorpdfstring{$\alg{sl}(2)$}{sl(2)}.}

From the expressions above we see that $\set{E_1,F_1, H_1}$ and 
$\set{E_3,F_3,H_3}$ generate the Hopf subalgebras $\env_{\hbar s_1}(\alg{sl}(2))$ and
$\env_{\hbar s_3}(\alg{sl}(2))$ respectively, deforming two of the $\alg{sl}(2)$
subalgebras of $\alg{d}(2,1;\epsilon)$. 
The final $\alg{sl}(2)$ subalgebra is
generated by the following combinations $\set{E_\comp,F_\comp,H_\comp}$
of the two even non-simple generators
and the Cartan generators  
\begin{align}
\label{eq:bosnonsimp1}
E_\comp & \eqq \frac{q-q^{-1}}{q_\comp^{-1} - q_\comp}[E_{32},E_{12}]_{-\hbar s_2}, 
\\ 
F_\comp   & \eqq \frac{q-q^{-1}}{q_\comp - q_\comp^{-1}} [F_{21},F_{23}]_{\hbar s_2},
\\
s_2 H_\comp & \eqq s_{1}H_{1} - 2  H_{2}+s_{3}H_{3},
\label{eq:bosnonsimp3}
\end{align}
where we introduced
\[
q_\comp\eqq e^{s_2 \hbar}.
\]
Their commutation relations are
\begin{align}
\comm{H_\comp}{E_\comp} & = 2 E_\comp, 
&\comm{H_\comp}{F_\comp} &= - 2 F_\comp,
&\comm{E_\comp}{F_\comp} &=
\frac{q_\comp^{H_\comp}-q_\comp^{- H_\comp}}{q_\comp-q_\comp^{-1}},
\end{align}
which are those of the Hopf algebra $\env_{\hbar s_2}(\alg{sl}(2))$.
However, as $E_\comp$ and $F_\comp$ are
non-simple generators their coproduct has a more complicated form. Indeed, as
a consequence of the requirement of compatibility between the coalgebra and algebra, the
coproduct for $\set{E_\comp,F_\comp,H_\comp}$ is
\begin{align}
\copro E_\comp 
& =E_\comp\otimes 1 + q_\comp^{-H_\comp}\otimes E_\comp
\nln&\quad 
-(q-q^{-1})q_\comp^{-1} E_{32}q_1^{-H_{1}}q_2^{- H_{2}}\otimes E_{12}
\nln&\quad 
+(q-q^{-1})q_\comp^{-1}\brk!{q_1 E_{132}+(q_1^{2}-1)E_{32}E_{1}}q_2^{-H_{2}}\otimes E_{2}
\nln&\quad 
+(q-q^{-1})q_\comp^{-1}(q_3^{2}-1)E_{3}q_1^{-H_{1}}q_2^{-2 H_{2}}\otimes E_{2}E_{12},
\\
\copro  F_\comp   
& = F_\comp  \otimes q_\comp^{ H_\comp}+ 1\otimes F_\comp  
\nln&\quad 
-(q-q^{-1})q_\comp F_{21}\otimes q_1^{H_{1}}q_2^{ H_{2}}F_{23}
\nln&\quad 
+(q-q^{-1})q_\comp F_{2}\otimes q_2^{H_{2}} 
\brk!{q_1^{-1}F_{213}+(q_1^{-2}-1)F_{1}F_{23}}
\nln&\quad
+(q-q^{-1})q_\comp(q_3^{-2}-1) F_{21}F_{2}\otimes q_1^{H_{1}}q_2^{2  H_{2}}F_{3},
\\
\copro H_\comp & = H_\comp \otimes 1 + 1 \otimes H_\comp,
\end{align}
where we see that $\copro E_\comp$ and $\copro F_\comp$ pick up a tail
involving the odd generators.

\medskip

It is this  deformed $\env_{\hbar s_2}(\alg{sl}(2))$ that will 
replace one $\alg{sl}(2)$ of \secref{sec:sl2+sl2} for the purpose 
of taking the contraction limit. Therefore, we give
the commutation relations of $\set{E_\comp,F_\comp,H_\comp}$ 
with the simple roots for convenience
\begin{align}
\comm{H_\comp}{E_i} & = \delta_{i2} E_i, &
\comm{H_\comp}{F_i} & = - \delta_{i2} F_i, 
\\
\comm{E_\comp}{E_{1}}&=0,
&\comm{E_\comp}{F_{1}}&=0,
\\
\comm{E_\comp}{E_{2}}&=(q_\comp^{-1}-1)E_{2}E_\comp,
&\comm{E_\comp}{F_{2}}&=q_1 q_\comp^{-1} \brk!{ E_{132}+\brk{q_1-q_1^{-1}}E_{32}E_{1} } q_2^{-H_{2}},
\\
\comm{E_\comp}{E_{3}}&=q_1(q-q^{-1})E_{32}E_{132},
&\comm{E_\comp}{F_{3}}&=q_1(q-q^{-1})E_{2}E_{12}q_3^{H_{3}},
\\
\comm{ F_\comp  }{F_1}&=0,
&\comm{ F_\comp  }{E_1}&=0,
\\
\comm{ F_\comp  }{F_2}&=(1-q_\comp) F_\comp  F_2,
&\comm{ F_\comp  }{E_2}&=q_1^{-1} q_\comp q_2^{ H_2} \brk!{ F_{213}+\brk{q_1^{-1}-q_1} F_1 F_{23} },
\\
\comm{ F_\comp  }{F_3}&=-q_1^{-1}(q-q^{-1})F_{213}F_{23},
&\comm{ F_\comp  }{E_3}&=-q_1^{-1}(q-q^{-1})q_3^{-H_3}F_{21}F_2.
\end{align}
Notice that these commutation relations as well as the coproduct of $E_\aut$
and $F_\aut$ do not exhibit a symmetry between the indices $1$ and $3$. This is
an artifact of the choice made in defining the even non-simple generators
\eqref{eq:bosnonsimp1,-,eq:bosnonsimp3}. A definition in terms of a
symmetric commutator $\comm{E_{12}}{E_{32}}$, however, leads to an
inconvenient basis for the purpose of presenting the R-matrix.

\paragraph{R-matrix.}

The R-matrix of $\env_\hbar(\alg{d}(2,1;\epsilon))$ was explicitly calculated in \cite{Thys:2001}.
The expression for the universal R-matrix depends on a choice of PBW basis for the positive and negative Borel subalgebras.
Adapted to our choice of basis for the positive Borel subalgebra
$\set{E_{2}^{n_2}E_{12}^{n_{12}}E_\comp^{n_\comp}E_{32}^{n_{32}}E_{132}^{n_{132}}
E_{1}^{n_1}E_{3}^{n_3}H_{1}^{m_1}H_{2}^{m_2}H_{3}^{m_3}|n_i,m_j\in \Natural_0}$
the R-matrix takes the form
\begin{align}
\label{eq:r-matd21a}
\rmat 
& =\exp\brk[s]!{-(q_2-q_2^{-1})E_{2}\otimes F_{2}}
\cdot\exp\brk[s]!{-(q_2-q_2^{-1})E_{12}\otimes F_{21}}
\nln & \quad
\cdot\exp_{-2 \hbar s_2} \brk[s]!{(q_\comp-q_\comp^{-1})E_\comp \otimes F_\comp  }
\nln & \quad
\cdot \exp\brk[s]!{-(q_2-q_2^{-1})E_{32}\otimes F_{32}}
\cdot\exp\brk[s]!{-(q_2-q_2^{-1})E_{132}\otimes F_{132}}
\nln & \quad
\cdot \exp_{-2\hbar s_1}\brk[s]!{(q_1-q_1^{-1})E_{1}\otimes F_{1}}
\cdot\exp_{-2\hbar s_3}\brk[s]!{(q_3-q_3^{-1})E_{3}\otimes F_{3}}
\nln & \quad
\cdot \exp\brk[s]!{\half\hbar (s_1 H_1 \otimes H_1+s_2 H_\comp \otimes H_\comp +s_3 H_3 \otimes H_3)}.
\end{align}

\subsection{Contraction limit}
\label{ssec:limit}

We will now apply the contraction limit of \secref{sec:sl2+sl2} with the role of
the generators $\set{E,F,H}$ of $\env_{\epsilon \hbar}(\alg{sl}(2))$ played by the
generators $\set{E_\comp,F_\comp,H_\comp}$ of $\env_\hbar(\alg{d}(2,1;\epsilon))$.
Indeed, upon using the parameterisation \eqref{eq:s123param} we find that
the commutation relations of the generators $\set{E_\comp,F_\comp,H_\comp}$ are
exactly those of $\env_{\epsilon \hbar}(\alg{sl}(2))$, 
while the coproduct for $E_\comp$ and $F_\comp$ now possesses a tail.
Our starting point is therefore the Hopf algebra $\env_\hbar (\alg{d}
(2,1;\epsilon))\otimes\env_{\tilde{\epsilon} \hbar}(\alg{sl}(2))$,
where we denote the generators of $\env_{\tilde{\epsilon} \hbar}
(\alg{sl}(2))$ by $\set{\tilde{E},\tilde{F},\tilde{H}}$ 
and where $\tilde \epsilon$ and $\epsilon$ are related in the same way 
as in \secref{sec:sl2+sl2}
\[
\tilde{\epsilon} (\epsilon)=-\epsilon +\xi \epsilon^2 +
\Order(\epsilon^3).
\] 
In particular the minus sign in 
the linear term is again required to have a well-defined, divergence-free
limit. Furthermore, the generators that we keep finite in the limit
are directly analogous to those of \secref{sec:sl2+sl2}
\begin{align}\label{eq:fg1}
 E_\aut&\eqq E_\comp+\tilde{E},
& E_\ctr&\eqq \epsilon E_\comp,\\
 F_\aut&\eqq  F_\comp  +\tilde{F},
& F_\ctr&\eqq \epsilon   F_\comp,  \\
\label{eq:fg2}
H_\aut&\eqq H_\comp+\tilde{H},
&  H_\ctr&\eqq\epsilon H_\comp.
\end{align}
The simple generators $\set{E_i,F_i,H_i}$,
and hence also the six odd non-simple generators \eqref{eq:fermns1,-,eq:fermns2}, 
all remain finite in the contraction limit $\epsilon \to 0$ ($s_1 \to 1$, $s_3 \to -1$).
This is consistent with the scaling of $ H_\ctr$, $E_\ctr$ and $F_\ctr$ with $\epsilon$ in
\eqref{eq:fg1,-,eq:fg2}. The former then generate the $\alg{sl}(2|2)$ part of
the maximally extended $\alg{sl}(2|2)$ Hopf algebra. In particular the Cartan matrix and
q-exponents all have a finite and non-degenerate
$\epsilon \to 0$ limit, and become those of $\alg{sl}(2|2)$
\begin{align}
a_{ij} & =
\begin{pmatrix}+ 2 & -1 & 0\\
+1 & 0 & -1\\
0 & -1 & +2
\end{pmatrix}_{ij} , 
& d_{i} & =\begin{pmatrix}1 & -1 & -1\end{pmatrix}_i.
\end{align}
Furthermore, the Serre relations \eqref{eq:serre1,-,eq:serre2}
reduce to the standard ones of $\env_\hbar(\alg{sl}(2|2))$.
On the other hand, as expected, the non-standard Serre elements,
$[E_{32},E_{12}]$ and $[F_{21},F_{23}]$, do not vanish in the 
contraction limit. Instead they become the generators $E_\ctr$ and $F_\ctr$ of the
extended algebra $\env_\hbar(\alg{psl}(2|2)\ltimes \Complex^3)$. 
Indeed, in the contracted algebra, the generators
$\set{E_\ctr,F_\ctr,H_\ctr}$ are related to the $\alg{sl}(2|2)$ generators
as
\begin{align}
E_\ctr &= -\frac{q-q^{-1}}{2\hbar} [E_{32},E_{12}], 
\\
F_\ctr &=\frac{q-q^{-1}}{2\hbar} [F_{21},F_{23}],
\\
H_\ctr &=  H_1 - 2 H_2 - H_3.
\end{align}
It now remains to confirm that the commutation relations involving the
generators \eqref{eq:fg1,-,eq:fg2} have a finite $\epsilon \to 0$ limit. After taking
the contraction limit the commutation relations of the generators
\eqref{eq:fg1,-,eq:fg2} with themselves are the same as in \secref{sec:sl2+sl2} and are given in
\eqref{eq:bosalg1,-,eq:bosalg5}. The commutation relations of the
generators \eqref{eq:fg1,-,eq:fg2} with the simple generators are 
such that $\set{E_\ctr, F_\ctr,H_\ctr}$ commute with them  
\begin{align}
\comm{  H_\ctr}{E_i}&=0, 
&\comm{  H_\ctr}{F_i}&=0,
\\
\comm{ E_\ctr}{H_{1,3}}& = \comm{ E_\ctr}{E_i} =\comm{ E_\ctr}{F_i} = 0,
& \comm{ F_\ctr}{H_{1,3}}& = \comm{ F_\ctr}{E_i} =\comm{ F_\ctr}{F_i} = 0,
\end{align}
while the generators $\set{E_\aut, F_\aut, H_\aut}$ have the following commutators
\begin{align}
\comm{ H_\aut}{E_i}&=\delta_{i2} E_i,
&\comm{ H_\aut}{F_i}&= -\delta_{i2} F_i,
\\
 \comm{ E_\aut}{E_1} &= \comm{ E_\aut}{F_1} = \comm{ E_\aut}{H_{1,3}} = 0,
&
\comm{ F_\aut}{E_1} &= \comm{ F_\aut}{F_1} =\comm{ F_\aut}{H_{1,3}}= 0 ,
\\
\comm{ E_\aut}{E_{2}}&=-\hbar E_{2} E_\ctr,
&\comm{ E_\aut}{F_{2}}&=q \brk!{E_{132}+(q-q^{-1})E_{32}E_{1}}q_2^{- H_{2}},
\\
\comm{ E_\aut}{E_{3}}&=(q-q^{-1})q E_{32}E_{132},
&
\comm{ E_\aut}{F_{3}}&=(q-q^{-1})q E_{2}E_{12}q_3^{ H_{3}},
\\
\comm{ F_\aut}{F_2}&=-\hbar F_\ctr F_2,
&
\comm{ F_\aut}{E_2}&=q^{-1}q_2^{ H_2}\brk!{F_{213}-\brk{q-q^{-1 }}F_1 F_{23}},
\\
\comm{ F_\aut}{F_3}&=-(q-q^{-1})q^{-1}F_{213}F_{23},
&
\comm{ F_\aut}{E_3}&=-(q-q^{-1})q^{-1}q_3^{- H_3}F_{21}F_2.
\end{align}
By repeated application of these relations one can then easily find the
commutation relations of the generators \eqref{eq:fg1,-,eq:fg2} with
the non-simple odd generators \eqref{eq:fermns1,-,eq:fermns2}.

The contraction limit of the coproduct for $E_\aut$ and $F_\aut$ is
\begin{align}
\copro E_\aut 
& = E_\aut\otimes1 +q^{-H_\ctr}\otimes E_\aut
-\hbar \brk{H_\aut+ \xi H_\ctr}q^{-H_\ctr}\otimes E_\ctr
\nln&\quad 
-(q-q^{-1})E_{32}q_1^{- H_{1}}q_2^{- H_{2}}\otimes E_{12}
\nln&\quad
 +(q-q^{-1})q\brk!{E_{132}+(q-q^{-1}) E_{32}E_{1}}q_2^{- H_{2}}\otimes E_{2}
\nln&\quad
-(q-q^{-1})^2 q^{-1} E_{3}q_1^{- H_{1}}q_2^{-2 H_{2}}\otimes E_{2}E_{12},
\\
\copro  F_\aut 
& = F_\aut\otimes q^{H_\ctr}+ 1\otimes F_\aut
+\hbar F_\ctr\otimes q^{H_\ctr}\brk{H_\aut+ \xi H_\ctr} 
\nln&\quad
-(q-q^{-1})F_{21}\otimes q_1^{ H_{1}}q_2^{ H_{2}}F_{23}
\nln&\quad 
+(q-q^{-1})q^{-1} F_{2}\otimes q_2^{ H_{2}}\brk!{F_{213}-(q-q^{-1})F_{1}F_{23}}
\nln&\quad
+(q-q^{-1})^2 q F_{21}F_{2}\otimes q_1^{ H_{1}}q_2^{2  H_{2}}F_{3},
\end{align}
while for $E_\ctr$ and $F_\ctr$ it is given in \eqref{eq:boscop3,-,eq:boscop4}
and is trivial for $H_\aut$ and $ H_\ctr$. The coproduct for the remaining generators
of $\env_\hbar(\alg{d}(2,1;\epsilon))$ does not depend on $E_\comp$ or
$F_\comp$ and hence remains unchanged in the limit up to setting $(s_1,s_2,s_3)=(1,0,-1)$.
Finally it is worth mentioning that the q-deformation of the $\alg{psl}(2|2)$-part
is still in place ($q\not\approx 1$) after the limit, while the q-deformation of the 
Poincar\'e part, as already seen in \secref{sec:sl2+sl2}, is mostly 
gone ($q\approx 1$) or reduced to $\hbar$ for the generators $\set{E_\aut,H_\aut,F_\aut}$.

\paragraph{R-matrix.} 

The R-matrix of $\env_\hbar (\alg{d}(2,1;\epsilon))\otimes\env_{\tilde{\epsilon} \hbar}(\alg{sl}(2))$ 
is given by the product of the individual R-matrices \eqref{eq:r-matd21a} and \eqref{eq:r-matsl2}.
The terms involving $\set{E_\ctr,F_\ctr,H_\ctr}$ and $\set{E_\aut,F_\aut,H_\aut}$ were already 
calculated in \secref{sec:sl2+sl2}.
The $\epsilon \to 0$ limit of the remaining terms is straight-forward 
and the complete R-matrix is given by
\begin{align}
\rmat 
& =\exp\brk[s]!{(q-q^{-1})E_{2}\otimes F_{2}}
\cdot \exp\brk[s]!{(q-q^{-1})E_{12}\otimes F_{21}}
\nln & \quad
\cdot\exp\brk[s]*{ 
-\frac{\xi}{2\hbar }\Li_{2}\brk!{4\hbar^2 E_\ctr\otimes F_\ctr} 
-\frac{\xi}{\hbar}\log\brk!{1 -4\hbar^2 E_\ctr\otimes F_\ctr}
}
\nln & \quad
\cdot\exp\brk[s]*{ -\frac{1}{2\hbar}
\brk!{ E_\ctr\otimes F_\aut +E_\aut\otimes F_\ctr}
\frac{\log\brk!{1 -4\hbar^2 E_\ctr\otimes F_\ctr}}{E_\ctr\otimes F_\ctr}
 }
\nln & \quad
\cdot \exp\brk[s]!{(q-q^{-1})E_{32}\otimes F_{32}}
\cdot \exp\brk[s]!{(q-q^{-1})E_{132}\otimes F_{132}}
\nln & \quad
\cdot \exp_{-2\hbar}\brk[s]!{(q-q^{-1})E_{1}\otimes F_{1}}
\cdot \exp_{2\hbar}\brk[s]!{(q^{-1}-q )E_{3}\otimes F_{3}}
\nln & \quad
\cdot \exp\brk[s]*{\half \hbar 
\brk{ H_1 \otimes H_1 
-  H_3 \otimes H_3 +   H_\ctr\otimes H_\aut + H_\aut\otimes  H_\ctr+ \xi H_\ctr\otimes H_\ctr}
}.
\end{align}

\paragraph{Identification with maximally extended \texorpdfstring{$\alg{sl}(2|2)$}{sl(2|2)}.}

Comparing the Hopf algebra and R-matrix found in the contraction 
limit with the results in \cite{Beisert:2016qei} we see
that we recover all relations upon identifying the generators 
used in \cite{Beisert:2016qei} as follows
\begin{align}
L&= E_\aut+ \xi E_\ctr,
&M&=-F_\aut,
&H_\aut&=H_\aut,
\\
P &=-\frac{2\hbar}{q-q^{-1}}E_\ctr,
&K &=\frac{2\hbar}{q-q^{-1}}F_\ctr,
&C&=\half H_\ctr,
\end{align}
as well as the parameter of \cite{Beisert:2016qei} as
\[
\kappa=2\xi.
\]
In particular, the R-matrix is in perfect agreement 
with the appropriate terms in \cite{Beisert:2016qei} upon using this identification.

Therefore, as claimed, we have recovered the maximally extended $\alg{sl}(2|2)$
Hopf algebra $\env_{\hbar,\xi}(\alg{sl}(2) \ltimes \alg{psl}(2|2) \ltimes
\Complex^3)$ as a contraction limit of $\env_\hbar(\alg{d}(2,1;\epsilon)) \otimes
\env_{\tilde{\epsilon} \hbar} (\alg{sl}(2))$.

\paragraph{Two copies of \texorpdfstring{$\alg{d}(2,1;\epsilon)$}{d(2,1;eps)}.}

We have seen that we can extend the contraction limit of \secref{sec:sl2+sl2}
to a contraction limit of $\env_\hbar(\alg{d}(2,1;\epsilon)) \otimes
\env_{\tilde{\epsilon} \hbar} (\alg{sl}(2))$ by replacing the
$\env_{\epsilon \hbar}(\alg{sl}(2))$ of the former contraction by an
$\alg{sl}(2)$ subalgebra inside $\alg{d}(2,1;\epsilon)$. One may ask
if we can also promote the 
$\env_{\tilde \epsilon \hbar} (\alg{sl}(2))$ 
factor to $\env_{\tilde\hbar} (\alg{d}(2,1;\tilde \epsilon))$.
Indeed this is possible without any additional complication. The contraction limit
of $\env_\hbar(\alg{d}(2,1;\epsilon)) \otimes
\env_{\tilde\hbar}(\alg{d}(2,1;\tilde{\epsilon}))$ leads to
$\env_{\hbar,\tilde\hbar,\xi}(\alg{sl}(2)\ltimes \alg{psl}(2|2)^{\directsum 2} \ltimes
\Complex^3)$ where we now have two copies of $\alg{psl}(2|2)$ sharing the
generators $\set{E_\aut,F_\aut,H_\aut}$ and $\set{E_\ctr,F_\ctr,H_\ctr}$.
Note that the resulting algebra has two q-deformed $\alg{psl}(2|2)$-parts
with independent q-deformation parameters $q=e^\hbar$ and $\tilde q=e^{\tilde \hbar}$.
Together with $\xi$, this algebra therefore carries 3 deformation parameters.

\section{Conclusion}
\label{sec:concl}

In this paper we have investigated contractions of quasi-triangular Hopf algebras.
Our focus was the following three examples
\begin{align}\label{eq:cl1}
\env_{\epsilon \hbar} (\alg{sl}(2)) \otimes  \env_{\tilde \epsilon \hbar} (\alg{sl}(2)) 
\quad & \to \quad \kap_\xi(\alg{iso}(3)),
\\\label{eq:cl2}
\env_{\hbar}(\alg{d}(2,1;\epsilon)) \otimes  \env_{\tilde \epsilon \hbar} (\alg{sl}(2)) 
\quad & \to \quad \env_{\hbar,\xi}(\alg{sl}(2) \ltimes \alg{psl}(2|2) \ltimes \Complex^3 ),
\\\label{eq:cl3}
\env_{\hbar}(\alg{d}(2,1;\epsilon)) \otimes  \env_{\tilde \hbar} (\alg{d}(2,1;\tilde \epsilon)) 
\quad & \to \quad \env_{\hbar,\tilde \hbar,\xi}(\alg{sl}(2) \ltimes \alg{psl}(2|2)^{\directsum 2} \ltimes \Complex^3 ).
\end{align}
In each case the initial algebra is semi-simple
and the resulting algebra is non-simple.
Exploring the corresponding freedom in the R-matrix it transpired that
certain choices had a finite contraction limit. This allowed us to
construct universal R-matrices for the contracted Hopf algebras.

\bigskip

The contraction \eqref{eq:cl1}, discussed in \secref{sec:sl2+sl2} and
\secref{sec:kappoincare}, led to a new one-parameter deformation of the 3D
kappa-Poincar\'e algebra.  This latter algebra is well-known, underlying
the physics on a certain non-commutative version of Minkowski space, and hence
it would be interesting to interpret the new parameter $\xi$.  We have also
obtained explicitly the universal R-matrix for this Hopf algebra.

It would be useful to further explore connections between our results and
those in the literature. One example of this would be the infinite boost limit.
In the classical analogue of $\kap_0(\alg{iso}(3))$ we can
consider a limit in which the distinguished generator becomes null, $P_0 \to
P_0 + P_1$ and $L_0 \to L_0 + L_1$ such that $\ad_{P_0 + P_1}^2 = \ad_{L_0 + L_1}^2 = 0$.  
In this case the Casimir term in the classical r-matrix is subleading and hence the leading
anti-symmetric part solves the classical Yang--Baxter equation
in its own right. The resulting classical r-matrix is then of jordanian type
\cite{Ogievetsky,Kulish,Tolstoy} and the universal R-matrix is expected to
reduce to a twist, the explicit form of which has been constructed
\cite{Borowiec:2008uj,Borowiec:2014aqa}. It would be interesting to see if this
expression can be recovered on taking the corresponding limit of \eqref{eq:bosrmat}.

\bigskip

The contractions \eqref{eq:cl2} and \eqref{eq:cl3}, 
discussed in \secref{sec:sl2+d}, are particularly important as they lead to
the maximally extended $\alg{sl}(2|2)$ Hopf algebra 
and the R-matrix of \cite{Beisert:2016qei} in a more systematic manner. 
Indeed lifting these contractions to deformed
affine algebras may provide a route to constructing the universal R-matrix for
the maximally extended affine $\alg{sl}(2|2)$ Hopf algebra.

To this end it appears convenient to consider the effect
of choosing different $\alg{sl}(2)$ subalgebras of $\alg{d}(2,1;\epsilon)$
for the contraction of \secref{sec:sl2+sl2}. For example, rather than
picking the most non-simple root $E_\comp$, one may choose one of the simple
roots $E_1$ or $E_3$. One may also consider the alternative Dynkin diagram with
all fermionic nodes.  The resulting Hopf algebras, while appearing different,
should be related.  For this the isomorphism permuting the three $\alg{sl}(2)$
algebras of $\alg{d}(2,1;\epsilon)$, and its extension to the quantum algebra,
will be relevant.

\bigskip

As we have seen, the maximally extended $\alg{sl}(2|2)$ algebra is a
supersymmetric extension of 3D kappa-Poincar\'e. However, the classical limit
of the former is not an ordinary super-Poincar\'e algebra because the
$\alg{sl}(2) \directsum \alg{sl}(2)$ subalgebra of $\alg{sl}(2|2)$ does not have
the canonical form of an R-symmetry.
To find such an algebra one can take a further limit
\begin{align}
Q &\to \gamma^{-1} Q, 
&
(E_\ctr,F_\ctr,H_\ctr) &\to \gamma^{-2} (E_\ctr,F_\ctr,H_\ctr),
& 
\gamma \to 0,
\end{align}
where $Q$ represents all odd generators. In this case the $\alg{sl}(2)
\directsum \alg{sl}(2)$ algebra becomes a derivation 
and thus indeed has the form of an R-symmetry. 
One might also consider a contraction of
$\alg{d}(2,1;\epsilon)$ with two of its own $\alg{sl}(2)$ algebras forming the
3D Poincar\'e algebra in the limit.  
In this case the R-symmetry would just consist of a single $\alg{sl}(2)$ algebra.  
It would be interesting to see whether these limits
can be implemented in $\env_\hbar (\alg{d}(2,1;\epsilon))$ and its extensions,
and hence if one can find universal R-matrices for 3D super-Poincar\'e algebras.

\bigskip

Finally, it is well-known that one can consider contraction limits in
two-dimensional sigma models with suitable global isometries, for example, the
flat space limit of anti-de-Sitter space. Recently such contraction limits were
extended to sigma models with q-deformed symmetries \cite{Pachol:2015mfa}.  In
light of our results it may now be worthwhile exploring these limits in more
detail for cases in which the isometry algebra is not simple
\cite{Klimcik:2014bta,Hoare:2014oua}.

\pdfbookmark[1]{Acknowledgements}{ack}
\section*{Acknowledgements}

The authors would like to thank 
M.\ de Leeuw
\unskip,
A.\ Sfondrini
\unskip,
A.\ Torrielli
and 
H.\ Zhang
for interesting discussions. 
This work is partially supported by grant no.\ 615203 
from the European Research Council under the FP7
and by the Swiss National Science Foundation
through the NCCR SwissMAP.

\begin{bibtex}[\jobname]

@article{Beisert:2016qei,
      author         = "Beisert, Niklas and de Leeuw, Marius and Hecht, Reimar",
      title          = "{Maximally extended sl(2$/$2) as a quantum double}",
      journal        = "J. Phys.",
      volume         = "A49",
      year           = "2016",
      number         = "43",
      pages          = "434005",
      doi            = "10.1088/1751-8113/49/43/434005",
      eprint         = "1602.04988",
      archivePrefix  = "arXiv",
      primaryClass   = "math-ph",
      SLACcitation   = "
}

@article{Beisert:2010jr,
      author         = "Beisert, Niklas and others",
      title          = "{Review of AdS/CFT Integrability: An Overview}",
      journal        = "Lett. Math. Phys.",
      volume         = "99",
      year           = "2012",
      pages          = "3-32",
      doi            = "10.1007/s11005-011-0529-2",
      eprint         = "1012.3982",
      archivePrefix  = "arXiv",
      primaryClass   = "hep-th",
      SLACcitation   = "
}

@Book{HubbBook,
      author         = "Fabian H. L. Essler and Holger Frahm and Frank G{\"o}hmann 
                         and Andreas Kl{\"u}mper and Vladimir E. Korepin",
      title          = "{The One-Dimensional Hubbard Model}",
      publisher      = "Cambridge University Press",
      year           = "2005",
      address        = "Cambridge, UK",
      doi            = "10.2277/0521802628",
}

@article{Shastry,
      author         = "Shastry, B. Sriram",
      title          = "{Decorated star-triangle relations and exact integrability 
                         of the one-dimensional Hubbard model}",
      publisher      = "Kluwer Academic Publishers-Plenum Publishers",
      year           = "1988",
      issn           = "0022-4715",
      journal        = "J. Stat. Phys.",
      volume         = "50",
      number         = "1-2",
      doi            = "10.1007/BF01022987",
      keywords       = "One-dimensional Hubbard model; exactly integrable systems; 
                        star-triangle relations",
      pages          = "57-79",
      language       = "English"
}

@article{Shastry:1986zz,
      author         = "Shastry, B. Sriram",
      title          = "{Exact Integrability of the One-Dimensional Hubbard Model}",
      journal        = "Phys. Rev. Lett.",
      volume         = "56",
      year           = "1986",
      pages          = "2453-2455",
      doi            = "10.1103/PhysRevLett.56.2453",
      SLACcitation   = "
}

@article{Beisert:2005tm,
      author         = "Beisert, Niklas",
      title          = "{The su(2$/$2) dynamic S-matrix}",
      journal        = "Adv. Theor. Math. Phys.",
      volume         = "12",
      year           = "2008",
      pages          = "945-979",
      doi            = "10.4310/ATMP.2008.v12.n5.a1",
      eprint         = "hep-th/0511082",
      archivePrefix  = "arXiv",
      primaryClass   = "hep-th",
      reportNumber   = "PUTP-2181, NSF-KITP-05-92",
      SLACcitation   = "
}

@article{Beisert:2006qh,
      author         = "Beisert, Niklas",
      title          = "{The Analytic Bethe Ansatz for a 
                         Chain with Centrally Extended su(2$/$2) Symmetry}",
      journal        = "J. Stat. Mech.",
      volume         = "0701",
      year           = "2007",
      pages          = "P01017",
      doi            = "10.1088/1742-5468/2007/01/P01017",
      eprint         = "nlin/0610017",
      archivePrefix  = "arXiv",
      primaryClass   = "nlin.SI",
      reportNumber   = "AEI-2006-074, PUTP-2211",
      SLACcitation   = "
}

@article{Giller:1992xg,
      author         = "Giller, S. and Kosinski, P. and Majewski, M. 
                        and Maslanka, P. and Kunz, J.",
      title          = "{More about $q$-deformed Poincar{\'e} algebra}",
      journal        = "Phys. Lett.",
      volume         = "B286",
      year           = "1992",
      pages          = "57-62",
      doi            = "10.1016/0370-2693(92)90158-Z",
      reportNumber   = "KFT-UL-2-92",
      SLACcitation   = "
}

@article{Lukierski:1991pn,
      author         = "Lukierski, Jerzy and Ruegg, Henri and Nowicki, Anatol and Tolstoy, Valerii N.",
      title          = "{$q$-deformation of Poincar{\'e} algebra}",
      journal        = "Phys. Lett.",
      volume         = "B264",
      year           = "1991",
      pages          = "331-338",
      doi            = "10.1016/0370-2693(91)90358-W",
      reportNumber   = "UGVA-DPT-1991-02-710",
      SLACcitation   = "
}

@article{Lukierski:1992dt,
      author         = "Lukierski, Jerzy and Nowicki, Anatol and Ruegg, Henri",
      title          = "{New quantum Poincar{\'e} algebra and $\kappa$-deformed field theory}",
      journal        = "Phys. Lett.",
      volume         = "B293",
      year           = "1992",
      pages          = "344-352",
      doi            = "10.1016/0370-2693(92)90894-A",
      reportNumber   = "UGVA-DPT-1992-07-776, LPTB-92-04",
      SLACcitation   = "
}

@article{Twietmeyer:1991mj,
      author         = "Twietmeyer, Eric",
      title          = "{Real forms of $U_q(g)$}",
      journal        = "Lett. Math. Phys.",
      volume         = "24",
      year           = "1992",
      pages          = "49-58",
      doi            = "10.1007/BF00430002",
      reportNumber   = "PRINT-91-0503 (HARVARD)",
      SLACcitation   = "
}

@article{Stachura,
      author         = "Stachura, Piotr",
      title          = "{Poisson-Lie structures on Poincar{\'e} 
                         and Euclidean groups in three dimensions}",
      journal        = "J. Phys.",
      volume         = "A31",
      year           = "1997",
      number         = "19",
      pages          = "4555",
      doi            = "10.1088/0305-4470/31/19/018"
}

@article{Zakrzewski,
      author        = "Zakrzewski, S.",
      title         = "{Poisson Structures on the Poincar{\'e} Group}",
      journal       = "Commun. Math. Phys.",
      volume        = "185",
      year          = "1997",
      pages         = "285-311",
      doi           = "10.1007/s002200050091",
      eprint        = "q-alg/9602001",
      archivePrefix = "arXiv",
      primaryClass  = "q-alg"
}

@article{Klimcik:2014bta,
      author         = "Klim\v{c}ik, Ctirad",
      title          = "{Integrability of the bi-Yang-Baxter sigma-model}",
      journal        = "Lett. Math. Phys.",
      volume         = "104",
      year           = "2014",
      pages          = "1095-1106",
      doi            = "10.1007/s11005-014-0709-y",
      eprint         = "1402.2105",
      archivePrefix  = "arXiv",
      primaryClass   = "math-ph",
      SLACcitation   = "
}

@article{Hoare:2014oua,
      author         = "Hoare, Ben",
      title          = "{Towards a two-parameter $q$-deformation 
                         of AdS$_3$ $\times$ S$^3$ $\times$ M$^4$ superstrings}",
      journal        = "Nucl. Phys.",
      volume         = "B891",
      year           = "2015",
      pages          = "259-295",
      doi            = "10.1016/j.nuclphysb.2014.12.012",
      eprint         = "1411.1266",
      archivePrefix  = "arXiv",
      primaryClass   = "hep-th",
      reportNumber   = "HU-EP-14-44",
      SLACcitation   = "
}

@article{Pachol:2015mfa,
      author         = "Pacho{\l}, Anna and van Tongeren, Stijn J.",
      title          = "{Quantum deformations of the flat space superstring}",
      journal        = "Phys. Rev.",
      volume         = "D93",
      year           = "2016",
      pages          = "026008",
      doi            = "10.1103/PhysRevD.93.026008",
      eprint         = "1510.02389",
      archivePrefix  = "arXiv",
      primaryClass   = "hep-th",
      reportNumber   = "HU-EP-15-48, HU-MATH-15-13",
      SLACcitation   = "
}

@article{Kirillov:1994en,
      author         = "Kirillov, Anatol N.",
      title          = "{Dilogarithm identities}",
      journal        = "Prog. Theor. Phys. Suppl.",
      volume         = "118",
      year           = "1995",
      pages          = "61-142",
      doi            = "10.1143/PTPS.118.61",
      eprint         = "hep-th/9408113",
      archivePrefix  = "arXiv",
      primaryClass   = "hep-th",
      SLACcitation   = "
}

@article{Thys:2001,
      author         = "Thys, Henrik",
      title          = "{R-Matrice universelle pour $U_h(D(2,1,x))$ et invariant d'entrelacs associ\'e}",
      journal        = "",
      volume         = "",
      year           = "2001",
      pages          = "",
      doi            = "",
      eprint         = "math/0104110",
      archivePrefix  = "arXiv",
      primaryClass   = "math",
      SLACcitation   = ""
}

@article{Borowiec:2014aqa,
      author         = "Borowiec, Andrzej and Pacho{\l}, Anna",
      title          = "{$\kappa$-Deformations and Extended $\kappa$-Minkowski Spacetimes}",
      journal        = "SIGMA",
      volume         = "10",
      year           = "2014",
      pages          = "107",
      doi            = "10.3842/SIGMA.2014.107",
      eprint         = "1404.2916",
      archivePrefix  = "arXiv",
      primaryClass   = "math-ph",
      SLACcitation   = "
}

@article{Lukierski:1993wxa,
      author         = "Lukierski, Jerzy and Ruegg, Henri",
      title          = "{Quantum $\kappa$-Poincar{\'e} in any dimension}",
      journal        = "Phys. Lett.",
      volume         = "B329",
      year           = "1994",
      pages          = "189-194",
      doi            = "10.1016/0370-2693(94)90759-5",
      eprint         = "hep-th/9310117",
      archivePrefix  = "arXiv",
      primaryClass   = "hep-th",
      SLACcitation   = "
}

@article{Maslanka,
      author         = "Maslanka, P.",
      title          = "{The $n$-dimensional kappa-Poincar{\'e} algebra and group}",
      journal        = "J. Phys.",
      volume         = "A26",
      year           = "1993",
      pages          = "L1251",
      doi            = "10.1088/0305-4470/26/24/001",
}

@article{Celeghini:1990xx,
      author         = "Celeghini, Enrico and Giachetti, Riccardo and Sorace, Emanuele and Tarlini, Marco",
      title          = "{The three-dimensional Euclidean quantum group E(3)$_q$ and its R-matrix}",
      journal        = "J. Math. Phys.",
      volume         = "32",
      year           = "1991",
      pages          = "1159-1165",
      doi            = "10.1063/1.529312",
      reportNumber   = "PRINT-90-0691 (FLORENCE)",
      SLACcitation   = "
}

@article{Drinfeld:1985rx,
      author         = "Drinfel'd, V. G.",
      title          = "{Hopf algebras and the quantum Yang-Baxter equation}",
      journal        = "Sov. Math. Dokl.",
      volume         = "32",
      year           = "1985",
      pages          = "254-258",
      SLACcitation   = "
}

@article{Drinfeld:1986in,
     author    = "Drinfel'd, V. G.",
     title     = "Quantum groups",
     journal   = "J. Math. Sci.",
     volume    = "41",
     year      = "1988",
     pages     = "898",
     doi       = "10.1007/BF01247086",
     SLACcitation  = "
}

@article{Jimbo:1985zk,
      author         = "Jimbo, Michio",
      title          = "{A q difference analog of $U(g)$ and the Yang-Baxter equation}",
      journal        = "Lett. Math. Phys.",
      volume         = "10",
      year           = "1985",
      pages          = "63-69",
      doi            = "10.1007/BF00704588",
      SLACcitation   = "
}

@article{Jimbo:1985vd,
      author         = "Jimbo, Michio",
      title          = "{A q-analog of $U(gl(n+1))$, Hecke algebra and the Yang-Baxter equation}",
      journal        = "Lett. Math. Phys.",
      volume         = "11",
      year           = "1986",
      pages          = "247",
      doi            = "10.1007/BF00400222",
      reportNumber   = "RIMS-517",
      SLACcitation   = "
}

@article{Celeghini:1990bf,
      author         = "Celeghini, E. and Giachetti, R. and Sorace, E. and Tarlini, M.",
      title          = "{Three Dimensional Quantum Groups from Contraction of SU(2)$_q$}",
      journal        = "J. Math. Phys.",
      volume         = "31",
      year           = "1990",
      pages          = "2548-2551",
      doi            = "10.1063/1.529000",
      SLACcitation   = "
}

@article{Nahm:1977tg,
      author         = "Nahm, W.",
      title          = "{Supersymmetries and their Representations}",
      journal        = "Nucl. Phys.",
      volume         = "B135",
      year           = "1978",
      pages          = "149",
      doi            = "10.1016/0550-3213(78)90218-3",
      reportNumber   = "CERN-TH-2341",
      SLACcitation   = "
}

@article{Faddeev:1993rs,
      author         = "Faddeev, L. D. and Kashaev, R. M.",
      title          = "{Quantum Dilogarithm}",
      journal        = "Mod. Phys. Lett.",
      volume         = "A9",
      year           = "1994",
      pages          = "427-434",
      doi            = "10.1142/S0217732394000447",
      eprint         = "hep-th/9310070",
      archivePrefix  = "arXiv",
      primaryClass   = "hep-th",
      reportNumber   = "HU-TFT-93-56",
      SLACcitation   = "
}

@article{Borowiec:2008uj,
      author         = "Borowiec, Andrzej and Pacho{\l}, A.",
      title          = "{kappa-Minkowski spacetime as the result of Jordanian
                        twist deformation}",
      journal        = "Phys. Rev.",
      volume         = "D79",
      year           = "2009",
      pages          = "045012",
      doi            = "10.1103/PhysRevD.79.045012",
      eprint         = "0812.0576",
      archivePrefix  = "arXiv",
      primaryClass   = "math-ph",
      SLACcitation   = "
}

@article{Ogievetsky,
      author         = "Ogievetsky, O.",
      title          = "{Hopf Structures on the Borel subalgebra of sl(2)}",
      journal        = "Suppl. Rend. Circ. Mat. Palermo, II. Ser.",
      volume         = "37",
      year           = "1993",
      pages          = "185",
}

@article{Kulish,
      author         = "Kulish, P. P. and Lyakhovsky, V. D. and Mudrov, A. I.",
      title          = "{Extended Jordanian twists for Lie algebras}",
      journal        = "J. Math. Phys.",
      volume         = "40",
      year           = "1999",
      pages          = "4569",
      doi            = "10.1063/1.532987",
      eprint         = "math/9806014",
      archivePrefix  = "arXiv",
      primaryClass   = "math",
}

@article{Tolstoy,
      author         = "Tolstoy, V. N.",
      title          = "{Chains of extended Jordanian twists for Lie superalgebras}",
      year           = "2004",
      eprint         = "math/0402433",
      archivePrefix  = "arXiv",
      primaryClass   = "math",
}

@article{Borowiec:2009vb,
      author         = "Borowiec, A. and Pacho{\l}, A.",
      title          = "{Classical basis for kappa-Poincar{\'e} algebra 
                         and doubly special relativity theories}",
      journal        = "J. Phys.",
      volume         = "A43",
      year           = "2010",
      pages          = "045203",
      doi            = "10.1088/1751-8113/43/4/045203",
      eprint         = "0903.5251",
      archivePrefix  = "arXiv",
      primaryClass   = "hep-th",
      SLACcitation   = "
}

@article{Kulish:1989sv,
      author         = "Kulish, P. P. and Reshetikhin, N. {\relax Yu}.",
      title          = "{Universal R matrix of the quantum superalgebra osp(2$/$1)}",
      journal        = "Lett. Math. Phys.",
      volume         = "18",
      year           = "1989",
      pages          = "143-149",
      doi            = "10.1007/BF00401868",
      SLACcitation   = "
}

@article{Kulish:1988gr,
      author         = "Kulish, P. P.",
      title          = "{Quantum Superalgebra osp(2$/$1)}",
      journal        = "J. Sov. Math.",
      volume         = "54",
      year           = "1989",
      pages          = "923-930",
      doi            = "10.1007/BF01101123",
      reportNumber   = "RIMS-615",
      SLACcitation   = "
}

@article{Matsumoto:2008ww,
      author         = "Matsumoto, Takuya and Moriyama, Sanefumi",
      title          = "{An Exceptional Algebraic Origin of the AdS/CFT Yangian
                        Symmetry}",
      journal        = "JHEP",
      volume         = "04",
      year           = "2008",
      pages          = "022",
      doi            = "10.1088/1126-6708/2008/04/022",
      eprint         = "0803.1212",
      archivePrefix  = "arXiv",
      primaryClass   = "hep-th",
      SLACcitation   = "
}

@book{Chari:1994pz,
  title={A Guide to Quantum Groups},
  author={Chari, V. and Pressley, A.N.},
  isbn={9780521558846},
  lccn={94241658},
  url={https://books.google.dk/books?id=bn5GNkLfnsAC},
  year={1995},
  publisher={Cambridge University Press}
}

@article{Faddeev:1987ih,
      author         = "Faddeev, L. D. and Reshetikhin, N. {\relax Yu}. and
                        Takhtajan, L. A.",
      title          = "{Quantization of Lie Groups and Lie Algebras}",
      journal        = "Leningrad Math. J.",
      volume         = "1",
      year           = "1990",
      pages          = "193-225",
      reportNumber   = "LOMI-E-14-87",
      SLACcitation   = "
}

@article{Plefka:2006ze,
      author         = "Plefka, Jan and Spill, Fabian and Torrielli, Alessandro",
      title          = "{On the Hopf algebra structure of the AdS/CFT S-matrix}",
      journal        = "Phys. Rev.",
      volume         = "D74",
      year           = "2006",
      pages          = "066008",
      doi            = "10.1103/PhysRevD.74.066008",
      eprint         = "hep-th/0608038",
      archivePrefix  = "arXiv",
      primaryClass   = "hep-th",
      reportNumber   = "HU-EP-06-22",
      SLACcitation   = "
}

@article{Gomez:2006va,
      author         = "G{\'o}mez, Cesar and Hern{\'a}ndez, Rafael",
      title          = "{The Magnon kinematics of the AdS/CFT correspondence}",
      journal        = "JHEP",
      volume         = "11",
      year           = "2006",
      pages          = "021",
      doi            = "10.1088/1126-6708/2006/11/021",
      eprint         = "hep-th/0608029",
      archivePrefix  = "arXiv",
      primaryClass   = "hep-th",
      reportNumber   = "CERN-PH-TH-2006-140, IFT-UAM-CSIC-06-37",
      SLACcitation   = "
}

@article{Torrielli:2007mc,
      author         = "Torrielli, Alessandro",
      title          = "{Classical r-matrix of the su(2$/$2) SYM spin-chain}",
      journal        = "Phys. Rev.",
      volume         = "D75",
      year           = "2007",
      pages          = "105020",
      doi            = "10.1103/PhysRevD.75.105020",
      eprint         = "hep-th/0701281",
      archivePrefix  = "arXiv",
      primaryClass   = "hep-th",
      reportNumber   = "MIT-CTP-3809",
      SLACcitation   = "
}

@article{Beisert:2007ty,
      author         = "Beisert, Niklas and Spill, Fabian",
      title          = "{The Classical r-matrix of AdS/CFT and its Lie Bialgebra
                        Structure}",
      journal        = "Commun. Math. Phys.",
      volume         = "285",
      year           = "2009",
      pages          = "537-565",
      doi            = "10.1007/s00220-008-0578-2",
      eprint         = "0708.1762",
      archivePrefix  = "arXiv",
      primaryClass   = "hep-th",
      reportNumber   = "AEI-2007-116, HU-EP-07-31",
      SLACcitation   = "
}

@article{Matsumoto:2007rh,
      author         = "Matsumoto, Takuya and Moriyama, Sanefumi and Torrielli,
                        Alessandro",
      title          = "{A Secret Symmetry of the AdS/CFT S-matrix}",
      journal        = "JHEP",
      volume         = "09",
      year           = "2007",
      pages          = "099",
      doi            = "10.1088/1126-6708/2007/09/099",
      eprint         = "0708.1285",
      archivePrefix  = "arXiv",
      primaryClass   = "hep-th",
      reportNumber   = "MIT-CTP-3853",
      SLACcitation   = "
}

@article{Beisert:2008tw,
      author         = "Beisert, Niklas and Koroteev, Peter",
      title          = "{Quantum Deformations of the One-Dimensional Hubbard
                        Model}",
      journal        = "J. Phys.",
      volume         = "A41",
      year           = "2008",
      pages          = "255204",
      doi            = "10.1088/1751-8113/41/25/255204",
      eprint         = "0802.0777",
      archivePrefix  = "arXiv",
      primaryClass   = "hep-th",
      reportNumber   = "AEI-2008-003, ITEP-TH-06-08",
      SLACcitation   = "
}

@article{KowalskiGlikman:2004qa,
      author         = "Kowalski-Glikman, Jerzy",
      title          = "{Introduction to doubly special relativity}",
      journal        = "Lect. Notes Phys.",
      volume         = "669",
      year           = "2005",
      pages          = "131-159",
      doi            = "10.1007/11377306_5",
      eprint         = "hep-th/0405273",
      archivePrefix  = "arXiv",
      primaryClass   = "hep-th",
      SLACcitation   = "
}

@article{Pachol:2011tp,
      author         = "Pacho{\l}, Anna",
      title          = "{$\kappa$-Minkowski spacetime: Mathematical formalism and
                        applications in Planck scale physics}",
      school         = "Wroclaw U.",
      year           = "2011",
      eprint         = "1112.5366",
      archivePrefix  = "arXiv",
      primaryClass   = "math-ph",
      note           = "PhD thesis",
      SLACcitation   = "
}

@article{Faddeev:1996iy,
      author         = "Faddeev, L. D.",
      title          = "{How algebraic Bethe ansatz works for integrable model}",
      booktitle      = "{Relativistic gravitation and gravitational radiation.
                        Proceedings, School of Physics, Les Houches, France,
                        September 26--October 6, 1995}",
      year           = "1996",
      pages          = "149-219",
      eprint         = "hep-th/9605187",
      archivePrefix  = "arXiv",
      primaryClass   = "hep-th",
      SLACcitation   = "
}

@article{Takhtajan:1979iv,
      author         = "Takhtajan, L. A. and Faddeev, L. D.",
      title          = "{The Quantum method of the inverse problem and the
                        Heisenberg XYZ model}",
      journal        = "Russ. Math. Surveys",
      volume         = "34",
      year           = "1979",
      number         = "5",
      pages          = "11-68",
      note           = "[Usp. Mat. Nauk34,no.5,13(1979)]",
      SLACcitation   = "
}

@article{Faddeev:1979gh,
      author         = "Faddeev, L. D. and Sklyanin, E. K. and Takhtajan, L. A.",
      title          = "{The Quantum Inverse Problem Method. 1}",
      journal        = "Theor. Math. Phys.",
      volume         = "40",
      year           = "1980",
      pages          = "688-706",
      reportNumber   = "LOMI-P-1-79",
      SLACcitation   = "
}

@article{Faddeev:1985qu,
      author         = "Faddeev, L. D. and Reshetikhin, N. {\relax Yu}.",
      title          = "{Integrability of the Principal Chiral Field Model in
                        (1+1)-dimension}",
      journal        = "Annals Phys.",
      volume         = "167",
      year           = "1986",
      pages          = "227",
      doi            = "10.1016/0003-4916(86)90201-0",
      reportNumber   = "LOMI-E-2-85",
      SLACcitation   = "
}

@article{Faddeev:1981ft,
      author         = "Faddeev, L. D. and Takhtajan, L. A.",
      title          = "{Spectrum and scattering of excitations in the
                        one-dimensional isotropic Heisenberg model}",
      journal        = "J. Sov. Math.",
      volume         = "24",
      year           = "1984",
      pages          = "241-267",
      doi            = "10.1007/BF01087245",
      note           = "[Zap. Nauchn. Semin.109,134(1981)]",
      SLACcitation   = "
}

@article{Faddeev:1994nk,
      author         = "Faddeev, L. D.",
      title          = "{Algebraic aspects of Bethe Ansatz}",
      journal        = "Int. J. Mod. Phys.",
      volume         = "A10",
      year           = "1995",
      pages          = "1845-1878",
      doi            = "10.1142/S0217751X95000905",
      eprint         = "hep-th/9404013",
      archivePrefix  = "arXiv",
      primaryClass   = "hep-th",
      reportNumber   = "ITP-SB-94-11",
      SLACcitation   = "
}

@article{Faddeev:1959yc,
      author         = "Faddeev, L. D.",
      title          = "{The Inverse problem in the quantum theory of
                        scattering}",
      journal        = "J. Math. Phys.",
      volume         = "4",
      year           = "1963",
      pages          = "72-104",
      doi            = "10.1063/1.1703891",
      note           = "english translation of Usp. Mat. Nauk 14, 57 (1959)",
      SLACcitation   = "
}

@article{SigmaNonComm,
      author         = "Paolo Aschieri and Harald Grosse and Giovanni Landi and Szabo (ed.), R.",
      title          = "Special Issue on Noncommutative Spaces and Fields",
      url            = "http://www.emis.de/journals/SIGMA/noncommutative.html", 
      year           = "2010",
      note           = "special issue in SIGMA",
}

@article{SigmaSpaceTime,
      author         = "G. Fiore and T. R. Govindarajan and J. Kowalski-Glikman and P. Kulish and Lukierski (ed.), J.",
      title          = "Special Issue on Deformations of Space-Time and its Symmetries",
      url            = "https://www.emis.de/journals/SIGMA/space-time.html", 
      year           = "2015",
      note           = "special issue in SIGMA",
}

\end{bibtex}

\bibliographystyle{nb}
\bibliography{\jobname}

\begin{thebibliography}{10}
\providecommand{\href}[2]{#2}
\providecommand{\arxivref}[2]{\href{http://arxiv.org/abs/#1}{#2}}
\providecommand{\doiref}[2]{\href{http://dx.doi.org/#1}{#2}}
\providecommand{\nbbstauthor}[1]{#1}
\providecommand{\nbbstjournal}[1]{\textsf{#1}}
\providecommand{\nbbsttitle}[1]{\textit{#1}}
\providecommand{\nbbsturl}[1]{\texttt{#1}}
\providecommand{\nbbsteprint}[1]{\texttt{#1}}
\providecommand{\nbbststyle}{\raggedright\small\parskip0pt}
\nbbststyle

\bibitem{Faddeev:1959yc}
\nbbstauthor{L.~D.~Faddeev},
\nbbsttitle{``{The Inverse problem in the quantum theory of scattering}''},
\nbbstjournal{\doiref{10.1063/1.1703891}{J.~Math.~Phys.~4,~72~(1963)}},
english translation of Usp. Mat. Nauk 14, 57 (1959).

\bibitem{Faddeev:1979gh}
\nbbstauthor{L.~D.~Faddeev, E.~K.~Sklyanin and L.~A.~Takhtajan},
\nbbsttitle{``{The Quantum Inverse Problem Method. 1}''},
\nbbstjournal{Theor.~Math.~Phys.~40,~688~(1980)}.

\bibitem{Faddeev:1996iy}
\nbbstauthor{L.~D.~Faddeev},
\nbbsttitle{``{How algebraic Bethe ansatz works for integrable model}''},
\nbbsteprint{\arxivref{hep-th/9605187}{hep-th/9605187}},
in: \nbbsttitle{``{Relativistic gravitation and gravitational radiation.
  Proceedings, School of Physics, Les Houches, France, September 26--October 6,
  1995}''},
pp.~149-219.

\bibitem{Drinfeld:1985rx}
\nbbstauthor{V.~G.~Drinfel'd},
\nbbsttitle{``{Hopf algebras and the quantum Yang-Baxter equation}''},
\nbbstjournal{Sov.~Math.~Dokl.~32,~254~(1985)}.

\bibitem{Drinfeld:1986in}
\nbbstauthor{V.~G.~Drinfel'd},
\nbbsttitle{``Quantum groups''},
\nbbstjournal{\doiref{10.1007/BF01247086}{J.~Math.~Sci.~41,~898~(1988)}}.

\bibitem{Jimbo:1985zk}
\nbbstauthor{M.~Jimbo},
\nbbsttitle{``{A q difference analog of $U(g)$ and the Yang-Baxter
  equation}''},
\nbbstjournal{\doiref{10.1007/BF00704588}{Lett.~Math.~Phys.~10,~63~(1985)}}.

\bibitem{Jimbo:1985vd}
\nbbstauthor{M.~Jimbo},
\nbbsttitle{``{A q-analog of $U(gl(n+1))$, Hecke algebra and the Yang-Baxter
  equation}''},
\nbbstjournal{\doiref{10.1007/BF00400222}{Lett.~Math.~Phys.~11,~247~(1986)}}.

\bibitem{Faddeev:1987ih}
\nbbstauthor{L.~D.~Faddeev, N.~{\relax Yu}.~Reshetikhin and L.~A.~Takhtajan},
\nbbsttitle{``{Quantization of Lie Groups and Lie Algebras}''},
\nbbstjournal{Leningrad~Math.~J.~1,~193~(1990)}.

\bibitem{Chari:1994pz}
\nbbstauthor{V.~Chari and A.~Pressley},
\nbbsttitle{``A Guide to Quantum Groups''},
Cambridge University Press (1995).

\bibitem{Kulish:1988gr}
\nbbstauthor{P.~P.~Kulish},
\nbbsttitle{``{Quantum Superalgebra osp(2$/$1)}''},
\nbbstjournal{\doiref{10.1007/BF01101123}{J.~Sov.~Math.~54,~923~(1989)}}.

\bibitem{Kulish:1989sv}
\nbbstauthor{P.~P.~Kulish and N.~{\relax Yu}.~Reshetikhin},
\nbbsttitle{``{Universal R matrix of the quantum superalgebra osp(2$/$1)}''},
\nbbstjournal{\doiref{10.1007/BF00401868}{Lett.~Math.~Phys.~18,~143~(1989)}}.

\bibitem{HubbBook}
\nbbstauthor{F.~H.~L.~Essler, H.~Frahm, F.~G{\"o}hmann, A.~Kl{\"u}mper and
  V.~E.~Korepin},
\nbbsttitle{``{The One-Dimensional Hubbard Model}''},
Cambridge University Press (2005),
Cambridge, UK.

\bibitem{Beisert:2010jr}
\nbbstauthor{N.~Beisert et~al.},
\nbbsttitle{``{Review of AdS/CFT Integrability: An Overview}''},
\nbbstjournal{\doiref{10.1007/s11005-011-0529-2}{Lett.~Math.~Phys.~99,~3~(2012)}},
\nbbsteprint{\arxivref{1012.3982}{arxiv:1012.3982}}.

\bibitem{Shastry:1986zz}
\nbbstauthor{B.~S.~Shastry},
\nbbsttitle{``{Exact Integrability of the One-Dimensional Hubbard Model}''},
\nbbstjournal{\doiref{10.1103/PhysRevLett.56.2453}{Phys.~Rev.~Lett.~56,~2453~(1986)}}.

\bibitem{Beisert:2005tm}
\nbbstauthor{N.~Beisert},
\nbbsttitle{``{The su(2$/$2) dynamic S-matrix}''},
\nbbstjournal{\doiref{10.4310/ATMP.2008.v12.n5.a1}{Adv.~Theor.~Math.~Phys.~12,~945~(2008)}},
\nbbsteprint{\arxivref{hep-th/0511082}{hep-th/0511082}}.

\bibitem{Beisert:2006qh}
\nbbstauthor{N.~Beisert},
\nbbsttitle{``{The Analytic Bethe Ansatz for a Chain with Centrally Extended
  su(2$/$2) Symmetry}''},
\nbbstjournal{\doiref{10.1088/1742-5468/2007/01/P01017}{J.~Stat.~Mech.~0701,~P01017~(2007)}},
\nbbsteprint{\arxivref{nlin/0610017}{nlin/0610017}}.

\bibitem{Gomez:2006va}
\nbbstauthor{C.~G{\'o}mez and R.~Hern{\'a}ndez},
\nbbsttitle{``{The Magnon kinematics of the AdS/CFT correspondence}''},
\nbbstjournal{\doiref{10.1088/1126-6708/2006/11/021}{JHEP~0611,~021~(2006)}},
\nbbsteprint{\arxivref{hep-th/0608029}{hep-th/0608029}}.

\bibitem{Plefka:2006ze}
\nbbstauthor{J.~Plefka, F.~Spill and A.~Torrielli},
\nbbsttitle{``{On the Hopf algebra structure of the AdS/CFT S-matrix}''},
\nbbstjournal{\doiref{10.1103/PhysRevD.74.066008}{Phys.~Rev.~D74,~066008~(2006)}},
\nbbsteprint{\arxivref{hep-th/0608038}{hep-th/0608038}}.

\bibitem{Torrielli:2007mc}
\nbbstauthor{A.~Torrielli},
\nbbsttitle{``{Classical r-matrix of the su(2$/$2) SYM spin-chain}''},
\nbbstjournal{\doiref{10.1103/PhysRevD.75.105020}{Phys.~Rev.~D75,~105020~(2007)}},
\nbbsteprint{\arxivref{hep-th/0701281}{hep-th/0701281}}.

\bibitem{Matsumoto:2007rh}
\nbbstauthor{T.~Matsumoto, S.~Moriyama and A.~Torrielli},
\nbbsttitle{``{A Secret Symmetry of the AdS/CFT S-matrix}''},
\nbbstjournal{\doiref{10.1088/1126-6708/2007/09/099}{JHEP~0709,~099~(2007)}},
\nbbsteprint{\arxivref{0708.1285}{arxiv:0708.1285}}.

\bibitem{Beisert:2007ty}
\nbbstauthor{N.~Beisert and F.~Spill},
\nbbsttitle{``{The Classical r-matrix of AdS/CFT and its Lie Bialgebra
  Structure}''},
\nbbstjournal{\doiref{10.1007/s00220-008-0578-2}{Commun.~Math.~Phys.~285,~537~(2009)}},
\nbbsteprint{\arxivref{0708.1762}{arxiv:0708.1762}}.

\bibitem{Nahm:1977tg}
\nbbstauthor{W.~Nahm},
\nbbsttitle{``{Supersymmetries and their Representations}''},
\nbbstjournal{\doiref{10.1016/0550-3213(78)90218-3}{Nucl.~Phys.~B135,~149~(1978)}}.

\bibitem{Beisert:2016qei}
\nbbstauthor{N.~Beisert, M.~de~Leeuw and R.~Hecht},
\nbbsttitle{``{Maximally extended sl(2$/$2) as a quantum double}''},
\nbbstjournal{\doiref{10.1088/1751-8113/49/43/434005}{J.~Phys.~A49,~434005~(2016)}},
\nbbsteprint{\arxivref{1602.04988}{arxiv:1602.04988}}.

\bibitem{Matsumoto:2008ww}
\nbbstauthor{T.~Matsumoto and S.~Moriyama},
\nbbsttitle{``{An Exceptional Algebraic Origin of the AdS/CFT Yangian
  Symmetry}''},
\nbbstjournal{\doiref{10.1088/1126-6708/2008/04/022}{JHEP~0804,~022~(2008)}},
\nbbsteprint{\arxivref{0803.1212}{arxiv:0803.1212}}.

\bibitem{KowalskiGlikman:2004qa}
\nbbstauthor{J.~Kowalski-Glikman},
\nbbsttitle{``{Introduction to doubly special relativity}''},
\nbbstjournal{\doiref{10.1007/11377306_5}{Lect.~Notes~Phys.~669,~131~(2005)}},
\nbbsteprint{\arxivref{hep-th/0405273}{hep-th/0405273}}.

\bibitem{SigmaNonComm}
\nbbstauthor{P.~Aschieri, H.~Grosse, G.~Landi and R.~Szabo~(ed.)},
\nbbsttitle{``Special Issue on Noncommutative Spaces and Fields''},
special issue in SIGMA,
\href{http://www.emis.de/journals/SIGMA/noncommutative.html}{\nbbsturl{http://www.emis.de/journals/SIGMA/noncommutative.html}}.

\bibitem{SigmaSpaceTime}
\nbbstauthor{G.~Fiore, T.~R.~Govindarajan, J.~Kowalski-Glikman, P.~Kulish and
  J.~Lukierski~(ed.)},
\nbbsttitle{``Special Issue on Deformations of Space-Time and its
  Symmetries''},
special issue in SIGMA,
\href{https://www.emis.de/journals/SIGMA/space-time.html}{\nbbsturl{https://www.emis.de/journals/SIGMA/space-time.html}}.

\bibitem{Pachol:2011tp}
\nbbstauthor{A.~Pacho{\l}},
\nbbsttitle{``{$\kappa$-Minkowski spacetime: Mathematical formalism and
  applications in Planck scale physics}''},
\nbbsteprint{\arxivref{1112.5366}{arxiv:1112.5366}},
PhD thesis.

\bibitem{Lukierski:1991pn}
\nbbstauthor{J.~Lukierski, H.~Ruegg, A.~Nowicki and V.~N.~Tolstoy},
\nbbsttitle{``{$q$-deformation of Poincar{\'e} algebra}''},
\nbbstjournal{\doiref{10.1016/0370-2693(91)90358-W}{Phys.~Lett.~B264,~331~(1991)}}.

\bibitem{Giller:1992xg}
\nbbstauthor{S.~Giller, P.~Kosinski, M.~Majewski, P.~Maslanka and J.~Kunz},
\nbbsttitle{``{More about $q$-deformed Poincar{\'e} algebra}''},
\nbbstjournal{\doiref{10.1016/0370-2693(92)90158-Z}{Phys.~Lett.~B286,~57~(1992)}}.

\bibitem{Lukierski:1993wxa}
\nbbstauthor{J.~Lukierski and H.~Ruegg},
\nbbsttitle{``{Quantum $\kappa$-Poincar{\'e} in any dimension}''},
\nbbstjournal{\doiref{10.1016/0370-2693(94)90759-5}{Phys.~Lett.~B329,~189~(1994)}},
\nbbsteprint{\arxivref{hep-th/9310117}{hep-th/9310117}}.

\bibitem{Maslanka}
\nbbstauthor{P.~Maslanka},
\nbbsttitle{``{The $n$-dimensional kappa-Poincar{\'e} algebra and group}''},
\nbbstjournal{\doiref{10.1088/0305-4470/26/24/001}{J.~Phys.~A26,~L1251~(1993)}}.

\bibitem{Celeghini:1990bf}
\nbbstauthor{E.~Celeghini, R.~Giachetti, E.~Sorace and M.~Tarlini},
\nbbsttitle{``{Three Dimensional Quantum Groups from Contraction of
  SU(2)$_q$}''},
\nbbstjournal{\doiref{10.1063/1.529000}{J.~Math.~Phys.~31,~2548~(1990)}}.

\bibitem{Celeghini:1990xx}
\nbbstauthor{E.~Celeghini, R.~Giachetti, E.~Sorace and M.~Tarlini},
\nbbsttitle{``{The three-dimensional Euclidean quantum group E(3)$_q$ and its
  R-matrix}''},
\nbbstjournal{\doiref{10.1063/1.529312}{J.~Math.~Phys.~32,~1159~(1991)}}.

\bibitem{Lukierski:1992dt}
\nbbstauthor{J.~Lukierski, A.~Nowicki and H.~Ruegg},
\nbbsttitle{``{New quantum Poincar{\'e} algebra and $\kappa$-deformed field
  theory}''},
\nbbstjournal{\doiref{10.1016/0370-2693(92)90894-A}{Phys.~Lett.~B293,~344~(1992)}}.

\bibitem{Faddeev:1993rs}
\nbbstauthor{L.~D.~Faddeev and R.~M.~Kashaev},
\nbbsttitle{``{Quantum Dilogarithm}''},
\nbbstjournal{\doiref{10.1142/S0217732394000447}{Mod.~Phys.~Lett.~A9,~427~(1994)}},
\nbbsteprint{\arxivref{hep-th/9310070}{hep-th/9310070}}.

\bibitem{Kirillov:1994en}
\nbbstauthor{A.~N.~Kirillov},
\nbbsttitle{``{Dilogarithm identities}''},
\nbbstjournal{\doiref{10.1143/PTPS.118.61}{Prog.~Theor.~Phys.~Suppl.~118,~61~(1995)}},
\nbbsteprint{\arxivref{hep-th/9408113}{hep-th/9408113}}.

\bibitem{Borowiec:2009vb}
\nbbstauthor{A.~Borowiec and A.~Pacho{\l}},
\nbbsttitle{``{Classical basis for kappa-Poincar{\'e} algebra and doubly
  special relativity theories}''},
\nbbstjournal{\doiref{10.1088/1751-8113/43/4/045203}{J.~Phys.~A43,~045203~(2010)}},
\nbbsteprint{\arxivref{0903.5251}{arxiv:0903.5251}}.

\bibitem{Twietmeyer:1991mj}
\nbbstauthor{E.~Twietmeyer},
\nbbsttitle{``{Real forms of $U_q(g)$}''},
\nbbstjournal{\doiref{10.1007/BF00430002}{Lett.~Math.~Phys.~24,~49~(1992)}}.

\bibitem{Stachura}
\nbbstauthor{P.~Stachura},
\nbbsttitle{``{Poisson-Lie structures on Poincar{\'e} and Euclidean groups in
  three dimensions}''},
\nbbstjournal{\doiref{10.1088/0305-4470/31/19/018}{J.~Phys.~A31,~4555~(1997)}}.

\bibitem{Zakrzewski}
\nbbstauthor{S.~Zakrzewski},
\nbbsttitle{``{Poisson Structures on the Poincar{\'e} Group}''},
\nbbstjournal{\doiref{10.1007/s002200050091}{Commun.~Math.~Phys.~185,~285~(1997)}},
\nbbsteprint{\arxivref{q-alg/9602001}{q-alg/9602001}}.

\bibitem{Beisert:2008tw}
\nbbstauthor{N.~Beisert and P.~Koroteev},
\nbbsttitle{``{Quantum Deformations of the One-Dimensional Hubbard Model}''},
\nbbstjournal{\doiref{10.1088/1751-8113/41/25/255204}{J.~Phys.~A41,~255204~(2008)}},
\nbbsteprint{\arxivref{0802.0777}{arxiv:0802.0777}}.

\bibitem{Thys:2001}
\nbbstauthor{H.~Thys},
\nbbsttitle{``{R-Matrice universelle pour $U_h(D(2,1,x))$ et invariant
  d'entrelacs associ\'e}''},
\nbbsteprint{\arxivref{math/0104110}{math/0104110}}.

\bibitem{Ogievetsky}
\nbbstauthor{O.~Ogievetsky},
\nbbsttitle{``{Hopf Structures on the Borel subalgebra of sl(2)}''},
\nbbstjournal{Suppl.~Rend.~Circ.~Mat.~Palermo,~II.~Ser.~37,~185~(1993)}.

\bibitem{Kulish}
\nbbstauthor{P.~P.~Kulish, V.~D.~Lyakhovsky and A.~I.~Mudrov},
\nbbsttitle{``{Extended Jordanian twists for Lie algebras}''},
\nbbstjournal{\doiref{10.1063/1.532987}{J.~Math.~Phys.~40,~4569~(1999)}},
\nbbsteprint{\arxivref{math/9806014}{math/9806014}}.

\bibitem{Tolstoy}
\nbbstauthor{V.~N.~Tolstoy},
\nbbsttitle{``{Chains of extended Jordanian twists for Lie superalgebras}''},
\nbbsteprint{\arxivref{math/0402433}{math/0402433}}.

\bibitem{Borowiec:2008uj}
\nbbstauthor{A.~Borowiec and A.~Pacho{\l}},
\nbbsttitle{``{kappa-Minkowski spacetime as the result of Jordanian twist
  deformation}''},
\nbbstjournal{\doiref{10.1103/PhysRevD.79.045012}{Phys.~Rev.~D79,~045012~(2009)}},
\nbbsteprint{\arxivref{0812.0576}{arxiv:0812.0576}}.

\bibitem{Borowiec:2014aqa}
\nbbstauthor{A.~Borowiec and A.~Pacho{\l}},
\nbbsttitle{``{$\kappa$-Deformations and Extended $\kappa$-Minkowski
  Spacetimes}''},
\nbbstjournal{\doiref{10.3842/SIGMA.2014.107}{SIGMA~10,~107~(2014)}},
\nbbsteprint{\arxivref{1404.2916}{arxiv:1404.2916}}.

\bibitem{Pachol:2015mfa}
\nbbstauthor{A.~Pacho{\l} and S.~J.~van~Tongeren},
\nbbsttitle{``{Quantum deformations of the flat space superstring}''},
\nbbstjournal{\doiref{10.1103/PhysRevD.93.026008}{Phys.~Rev.~D93,~026008~(2016)}},
\nbbsteprint{\arxivref{1510.02389}{arxiv:1510.02389}}.

\bibitem{Klimcik:2014bta}
\nbbstauthor{C.~Klim\v{c}ik},
\nbbsttitle{``{Integrability of the bi-Yang-Baxter sigma-model}''},
\nbbstjournal{\doiref{10.1007/s11005-014-0709-y}{Lett.~Math.~Phys.~104,~1095~(2014)}},
\nbbsteprint{\arxivref{1402.2105}{arxiv:1402.2105}}.

\bibitem{Hoare:2014oua}
\nbbstauthor{B.~Hoare},
\nbbsttitle{``{Towards a two-parameter $q$-deformation of AdS$_3$ $\times$
  S$^3$ $\times$ M$^4$ superstrings}''},
\nbbstjournal{\doiref{10.1016/j.nuclphysb.2014.12.012}{Nucl.~Phys.~B891,~259~(2015)}},
\nbbsteprint{\arxivref{1411.1266}{arxiv:1411.1266}}.

\end{thebibliography}

\end{document}